\documentclass[10pt,preprintnumbers,nofootinbib]{revtex4-2}%
\usepackage[T1]{fontenc}
\usepackage[utf8]{inputenc}
\usepackage{mathtools}
\usepackage{amsmath}
\usepackage{amssymb}
\usepackage{graphicx}
\usepackage[colorlinks=true, pdfstartview=FitV, linkcolor=blue, citecolor=red, urlcolor=magenta]{hyperref}
\usepackage{amsfonts}
\usepackage{subfigure}
\usepackage{cleveref}
\usepackage{listings}
\usepackage{xcolor}
\usepackage{array}
\usepackage{url}
\usepackage{comment}
\usepackage{todonotes}
\usepackage{dsfont}
\usepackage{appendix}

\usepackage{tensor}

\usepackage{xcolor}
\usepackage{soul}
\allowdisplaybreaks

\foreach \x in {A,...,Z}{\expandafter\xdef\csname \x\endcsname{\noexpand\mathcal{\x}}} 
\foreach \x in {A,...,Z}{\expandafter\xdef\csname I\x\endcsname{\noexpand\mathbb{\x}}} 
\foreach \x in {A,...,Z,a,...,z}{\expandafter\xdef\csname frak\x\endcsname{\noexpand\mathfrak{\x}}}
\DeclareMathOperator{\dee}{d}

\DeclareMathOperator{\Tr}{Tr}

\graphicspath{{Figures/}}

\begin{document}

\title{Spectral torsion of the internal noncommutative geometry of the Standard Model
}

\author{Ludwik D\k{a}browski}
\email{dabrow@sissa.it}
\affiliation{Scuola Internazionale Superiore di Studi Avanzati, Via Bonomea 265, Trieste, 34136, Italy}

\author{Sugato Mukhopadhyay}
\email{smukhopa@sissa.it}
\affiliation{Scuola Internazionale Superiore di Studi Avanzati, Via Bonomea 265, Trieste, 34136, Italy}

\author{Filip Po\v{z}ar}\thanks{Corresponding author}
\email{filip.pozar@irb.hr}
\affiliation{Rudjer Bo\v{s}kovi\'c Institute, Bijeni\v cka  c.54, HR-10000 Zagreb, Croatia}

\begin{abstract}
We compute the nonvanishing spectral torsion functional of the internal part 
of the noncommutative geometry behind the Standard Model. We show that with a suitable modification of the usual differential graded calculus
it matches an analogous functional constructed in terms of the connection. We study also the impact of the torsion on the other spectral fuctionals, which correspond to geometric invariants such as volume integral, metric and Einstein tensors, and scalar curvature. We discuss the impact of the SM Yukawa couplings and the Majorana mass matrix on our results.
\end{abstract}
\maketitle
\section{Introduction}
From a physical perspective, noncommutative geometry (NCG) is well motivated and is regarded as a firm candidate for probing aspects of quantum gravity due to the natural implementation of noncommutativity of spacetime functions, which include coordinate functions. Moreover, NCG in the Connes approach was shown in \cite{Chamseddine:2006ep} to reproduce the Euclidean Standard Model's (SM) Lagrangian from very natural spectral data. Further, it potentially offers gripping explanations of some of the Standard Model's properties which are, so far, imposed by hand. Notably, the almost commutative spectral triple underpinning Standard Model generates both the correct gauge fields and the Higgs field (with the symmetry breaking mechanism) as the connection on the corresponding almost commutative manifold \cite{Chamseddine:2006ep}. Additionally, this spectral triple automatically correctly sets the representations of all of the fields in the Lagrangian without any room for choice \cite{Chamseddine:2006ep, Chamseddine:2007ia, Schucker:2010uz}. All of this indicates that NCG is important not only in physics beyond General Relativity but also beyond the Standard Model.\\

In this paper we thoroughly investigate the noncommutative geometry of the internal part\footnote{We believe it should be relevant also after a supposed passage to the Lorentzian signature.} of the spectral triple behind the Standard Model. More precisely, we analyze the Hilbert space $\mathcal{H}$ and the Dirac operator $D$ as in \cite{ConnesMarcolli2007}, which describe the exact contents of the matter fields and dynamics of the Standard Model. However, we complexify the algebra and its representation on $\mathcal{H}$ in order to relate the spectrally defined geometric tensors \cite{Dabrowski:2023irb} to the differential calculus defined at the algebraic level, as in literature so far (cf. \cite{beggs2020quantum}).

Our study is part of a general program of linking and unifying the spectral geometry with differential graded calculus. Corresponding to various pivotal geometric tensors, it boils down to equating certain spectrally and algebraically defined functionals of noncommutative differential forms. This method faces the underlying difficulty that while the spectral functionals are defined completely using the information encoded in the spectral triple itself, on the algebraic side of differential graded calculus it is the linear connection\footnote{And the choice of second order differential calculus} that plays the central role and is inputted as extra data. Marrying those two quite different approaches would mean, as was discussed in \cite{dkabrowski2024algebraic}, to find the extra algebraic data (ideally unique) which reproduces the spectral functionals identically. The results of \cite{dkabrowski2024algebraic} suggest that this marriage is possible for the case of Connes--Lott model of a two--sheeted space. In our paper we consider the analogous questions for the more involved but more relevant to physics spectral triple behind the Standard Model.

Out of pivotal geometric tensors, torsion is the most involved one which has descriptions both on algebraic and spectral side. However, recently in \cite{Bochniak:2024duf} it was demonstrated that classical torsion faces impediments to be adjoined to the standard (torsion free) General Relativity, if the Einstein tensor is going to be obtained  in spectral manner, using e.g. the Dirac operator. Thus in this paper we focus on the quantum/noncommutative torsion. It has been shown to nonnvanish already in the simplest case of quantum two--sheeted space \cite{dkabrowski2024algebraic}. In this respect, a physically very relevant instance is the noncommutative geometry \cite{Chamseddine:2006ep, Chamseddine:2007ia, Schucker:2010uz} behind the Standard Model of fundamental particles,  and as the first step we analyse here its (quantum) internal part. It is described by a certain finite noncommutative spectral triple, which, as our first main result of this paper, we show to correspond to a noncommutative manifold with definitely nonvanishing torsion. 
As our second main result we obtain that it is possible to equate this spectral torsion with the  algebraically defined torsion on the graded differential calculus, 
similarly to the case of doubly sheeted space, but now only after slightly relaxing one of conditions from \cite{mesland2025existence}.

The internal Standard Model's quantum torsion, that we have obtained in agreement in two different approaches, suggests a new and previously unexplored source of geometric structure that could have cosmological implications\footnote{After transitioning to the full almost commutative spectral triple, given as the product of the internal spectral triple and the (commutative) spin manifold's spectral triple.}. In particular, internal torsion modifies the Einstein and scalar curvature spectral functionals in a way that depends on the Standard Model’s Yukawa couplings. At high energies (such as those relevant in the early universe), these torsional contributions may feed back into effective gravitational dynamics, potentially acting as a small but nontrivial “quantum spin–geometry” correction. Given that torsion in other contexts has been shown to drive nonstandard cosmological behavior (e.g., in bounce or inflationary scenarios \cite{Poplawski:2010kb, Dzhunushaliev:2012vb}), our findings open a plausible bridge: the internal quantum torsion uncovered in the noncommutative-geometric formalism could seed modifications to cosmological evolution. This connection motivates further study of how torsion originating from noncommutative geometry might leave residual imprints in early-universe cosmology. Additionally, there are existing efforts at interpreting torsion as dark matter and dark energy \cite{Grensing:2020maa, delaCruzDombriz:2021nrg, Pereira:2022cmu}. It is especially appealing that our approach introduces torsion to spacetime without requiring any new matter fields.\\

As ancilliary results, we compute the noncommutative integral and the metric tensor functional (both of which are independent of torsion) and successfully equate them with their corresponding algebraic counterparts. 
Next, we also compute the functional of which the density yields Ricci scalar curvature (in presence of torsion) and find that it is nonzero. In doing this we encounter an interesting gap between spectral side \cite{Dabrowski:2023irb} and algebraic side \cite{mesland2025existence}. Namely, we find that the spectral scalar curvature of the considered spectral triple necessarily depends on the Majorana mass term $\Upsilon_R$. On the other hand, regarding the algebraic side, the noncommutative 1--forms do not depend on the Majorana term $\Upsilon_R$ and neither does the metric or any other algebraic ingredients constructed so far in \cite{mesland2025existence}. This blindness of NC 1--forms to the Majorana mass is physically reflected in the fact that Majorana mass does not arise from a Higgs mechanism, which is completely encoded in the NC 1--forms of the finite spectral triple. Having said all that, our results point to the fact that if algebraic scalar curvature is defined similarly to other tensors in the algebraic approach, it would be impossible to equate algebraic and spectral scalar curvatures. Similar conclusions, although more weakly, are hinted also for the Einstein and Ricci tensors. This finding is important because it, firstly, suggests that the unification program
of \cite{dkabrowski2024algebraic} can only be achieved with an improved formalism of noncommutative geometry, although we are aware that
this is not an easy task. Secondly, in this example it appears that the gap in the unification program is intimately connected to the Majorana mass term in the NC Standard Model, which opens a very interesting question regarding its noncommutative geometric interpretation. We suspect that answering one of these points will necessarily increase understanding of the other.\\
\tableofcontents
\section{Preliminaries} \label{11sep25sm1}

A spectral triple is the data $(A,\H,D)$, where $A$ is a unital *--algebra, $\H$ is a Hilbert space admitting a (faithful) representation of $A$, and $D = D^\ast$ is a Dirac operator satisfying certain analytic conditions in general. However, the object of our study is a finite model, hence we shall not address these technicalities.

 For a unital *--algebra $A$, the universal differential calculus refers to the Differential Graded Algebra (DGA) $(\Omega^\bullet_u(A) = \oplus_{k=0}^\infty \Omega^k_u(A), \delta)$ defined over the degree zero algebra $\Omega^0_u = A$, where the exterior derivative $\delta$ satisfies the following Leibniz relation over degree zero elements $a$ and $b$:
\begin{equation}
    \delta(ab) = \delta(a) b + a \delta(b), \  a,b \in A.
\end{equation}

The space of degree one, $\Omega^1_u(A)$, is known as the space of universal 1--forms and as an $A$--bimodule is isomorphic the the kernel of the multiplication map $\mu:A\otimes_{\mathbb{C}} A \to A$ via the identification
\begin{equation}
    \sum_j a_j \otimes b_j \mapsto \sum_j a_j \delta(b_j).
\end{equation}
A general space $\Omega^k_u$ of universal $k$--forms, is given by finite sums
\begin{equation}
    \sum a_0 \delta(a_1) \dots \delta(a_k), \ a_0, a_1, \dots, a_k \in A.
\end{equation}
The exterior derivative $\delta$ is a degree one differential $\delta: \Omega^k_u \to \Omega^{k+1}_u$ satisfying the relations $\delta^2 = 0$,
\begin{equation}
\begin{aligned}
    \delta( \sum a_0 \delta(a_1) \dots \delta(a_k) ) =\sum \delta(a_0) \delta(a_1) \dots \delta(a_k),\\
    \delta(\omega \eta) = \delta(\omega) \eta + (-1)^{\text{deg}(\omega)} \omega \delta(\eta), \ \omega, \eta \in \Omega^\bullet_u.
    \end{aligned}
\end{equation}
Given a spectral triple $(A,H,D)$, the representation $\pi:A \to B(\H)$ extends to the space of universal 1--forms via the map $\pi_D:\delta(a) \mapsto [D, a]$. The image $\Omega^1_D = \pi_D(\Omega^1_u)$ together with the derivative map 
\begin{equation}
\dee_D(a) = [D, a]
\label{d_D}
\end{equation}
forms the first order differential calculus of the spectral triple. Thus,
\begin{equation}
        \Omega^1_D = \pi_D(\Omega^1_u) = \text{span}_\IC \left\{ \pi(a)[D,\pi(b)] : a,b \in A\right\}.
\label{Omega1_D}
\end{equation}
Since the representation $\pi$ is faithful, we identify the operator $\pi(a)$ with the algebra element $a$, and henceforth dispense with the symbol $\pi$.\\

Classically, for a differentiable manifold $M$, the algebra $A$ corresponds to the space of smooth complex functions $C^\infty(M,\IC)$, $\Omega^1_D$ is the space of covectors coming out of smooth sections $\Gamma^\infty(T^\ast(M)_\IC)$ of complexified cotangent bundle, and the exterior derivative $\dee$, acting on a smooth function $f$ is given in local coordinates by the formula
\[ \dee(f) = \partial_i (f) \dee \! x^i.\]
To achieve the results later in the paper, we need to define differential forms of the second order. Classically, this is the complex space $\Omega^2(M)_\IC$ which is spanned locally, as a module over $C^\infty(M)_\IC$, by the two forms $\dee \! x^i \wedge \dee \! x^j$ satisfying the antisymmetric relation
\[ \dee \! x^i \wedge \dee \! x^j = - \dee \! x^j \wedge \dee \! x^i. \]
We note here, that the space of two forms is isomorphic to the anti--symmetric part of the space of two tensors $\Omega^1(M)_\IC \otimes_{C^\infty(M)_\IC} \Omega^1(M)_\IC.$ The exterior derivative $\dee : C^\infty(M)_\IC \to \Omega^1(M)_\IC$ extends to the map
\[ \dee : \Omega^1(M)_\IC \to \Omega^2(M)_\IC, \quad \dee(f \dee( g)) = \dee(f) \wedge \dee(g). \]
Coming back to the noncommutative setup, we would like to realize the universal differential map
\begin{equation}
    \delta( a \delta(b)) \mapsto \delta(a) \delta(b)
\end{equation}
on operators using the representation $\pi_D$. However, the obstruction in this regard is the fact that the map
\begin{equation}
    a[D,b] \mapsto [D,a][D,b]
\end{equation}
need not be well--defined (due to the existence od divisors of $0$ in $B(\mathcal{H})$).\\
At this point we arrive at the first divergence of the algebraic formalisms behind this article. The two main contenders for defining an extension of the representation $\pi_D$ to second order universal forms are the approach of Connes (\cite{connes}) and the more recent formulation due to Mesland--Rennie (\cite{mesland2025existence}). We start with the formulation of Connes.

Let us denote by $J^2_D$ the vector space given as
\begin{equation} \label{16sep25sm1}
    J^2_D =  \left\{ \sum [D,a_j][D,b_j] : \sum a_j[D,b_j] = 0. \right\}
\end{equation}

This is precisely the obstruction to have a well--defined differential coming out of $\Omega^1_D$ and sometimes called the junk 2--forms. We thus quotient the bimodule $\pi_D(\Omega^2_u) = {}_A\text{span} \left\{ [D,a][D,b] \right\}$ by $J^2_D$. The proof that $J^2_D$ is a bimodule, making the quotient a well--defined bimodule, is relegated to \ref{7oct25sm1}.

Following Connes \cite{connes}, we define the bimodule
\begin{equation} \label{11sep25sm2}
    \Omega^2_D := \pi_D(\Omega^2_u) / J^2_D
\end{equation}
as the space of 2--forms of the spectral triple. Then $\delta$ descends to a well defined map from $\Omega^1_D$ to $\Omega^2_D$. The analogous construction can be done for higher order universal forms, but we shall not need them for this article. The main drawback of this construction is the fact that the quotient $\pi_D(\Omega^2_u) / J^2_D$, while a bimodule, consists of classes of operators, as opposed to genuine operators. In \cite{connes}, the space of classes of operators were identified with the orthogonal complement $\pi_D(\Omega^2_u) / J^2_D \cong \left(J^2_D\right)^\perp$ but we will show that this part of the construction is not appropriate for our purposes.\\

Next we recall the recent construction of Mesland--Rennie (\cite{mesland2025existence}), which gives us an alternative approach. But first, we define
\begin{equation}
    T^2_D := \Omega^1_D \otimes_A \Omega^1_D.
\end{equation}
Here, the balanced tensor product over the algebra is the quotient given by
\begin{equation}
    \Omega^1_D \otimes_A \Omega^1_D := (\Omega^1_D \otimes_\IC \Omega^1_D)/\text{span}_\IC \left\{\omega_1 a \otimes \omega_2 - \omega_1 \otimes a \omega_2 \right\},
\end{equation}
where $\omega_j$ are elements in $\Omega^1_D$ and $a$ is an arbitrary element in the algebra $A$. Similarly to how one can use the map $\pi_D$ to represent universal k--forms in $B(\mathcal{H})$
$$\pi_D (a_0 \delta(a_1)...\delta(a_k)) = a_0 \left[D,a_1\right]...\left[D,a_k\right]\;,$$
the map $\pi_D^{\otimes k}$ is a natural analogue for $k$--tensor forms with its definition given as
\begin{equation}
    \pi_D^{\otimes k} \left( a_0 \delta(a_1)...\delta(a_k)\right) = a_0\left[D,a_1\right]\otimes...\otimes \left[D,a_k\right] \in T^k_D\;.
\end{equation}
Also similarly to the previous analysis of Connes' k--forms, but now restricted to $k=2$ for our purposes, there can exist pairs of $a_j,b_j\in A$ giving rise to a universal 1--form
$$\omega = \sum_j a_j \delta(b_j)$$
such that $\pi_D(\omega) = 0$ but $\pi_D^{\otimes 2}(\delta(\omega)) \neq 0$. With that said, the "junk" 2--tensors are defined as
\begin{equation}
   JT^2_D := \left\{\pi_D^{\otimes 2}(\delta(\omega)) : \pi_D(\omega) = 0, \omega\in \Omega^1_u  \right\}\;,
\end{equation}
i.e., the exact 2--tensors which come from universal oneforms $\omega$ for which the representation $\pi_D$, which is faithful only on the algebra $A$, maps the nonzero universal 1--form $\omega$ to a zero operator on $B(\mathcal{H})$. From the definitions of $T^2_D$ and $\pi_D(\Omega^2_u)$, the following $\mathbb{C}$-linear multiplication map is surjective
\begin{equation}
    \begin{split}
        &\mu:T^2_D \rightarrow \pi_D(\Omega^2_u) \\
        &\mu(\omega_1 \otimes \omega_2) := \omega_1 \omega_2\;,\quad \omega_1,\omega_2 \in \pi_D (\Omega^1_u)\;.
    \end{split}
\end{equation}
Additionally, the space of 2--tensors $T^2_D$ is, by definition as a balanced tensor product of bimodules, a bimodule. A nontrivial result is that the junk 2--tensors (and all other junk $k$--tensors as well) are a bimodule as well. The proof of this is similar to the one in the \ref{7oct25sm1}. Following \cite{mesland2025existence}, we define the bimodule of tensor 2--forms as $\Lambda^2_D$ given by
\begin{equation} \label{11sep25sm3}
    \Lambda^2_D := T^2_D/JT^2_D.
\end{equation}
Given an idempotent bimodule map $\Psi$ on $T^2_D$ landing onto $JT^2_D$, we define the exterior derivative $\dee_\Psi : \Omega^1_D \to \Lambda^2_D$ by the formula
\begin{equation} \label{16sep25sm5}
    \dee_\Psi : a [D,b] \mapsto (1-\Psi) \left([D,a] \otimes [D,b]\right),
\end{equation}
and the data $(\Lambda^2_D, \dee_\Psi)$ define a second order differential calculus on the spectral triple. Classically, $JT^2_D$ corresponds to the symmetric part of the two tensors $\Omega^2(M)_\IC \otimes_{C^\infty(M)_\IC} \Omega^2(M)_\IC$, and the $\Psi$ map is locally given by
\[ \Psi: \dee \! x^i \otimes \dee \! x^j = \frac{1}{2}(\dee \! x^i \otimes \dee \! x^j + \dee \! x^j \otimes \dee \! x^i). \]
Classically, the kernel of the idempotent is thus the space of anti--symmetric two tensors, which as we recall, is isomorphic to the space two forms $\Omega^2(M)_\IC$.\\

The idempontent $\Psi$ provides a generalized way of avoiding working with classes of operators, compared to Connes' approach. We remind, in Connes' approach the space of 2--forms $\Omega^2_D$, which is given as the quotient $\Omega^2_D = \pi_D(\Omega^2_u)/J^2_D$, is identified with the orthogonal complement of $J^2_D$ inside $\pi_D(\Omega^2_u)$. As such, $\Omega^2_D$, which is defined as a quotient and by definition is a set of classes of operators in $B(\mathcal{H})$, is identified with concrete operators. Using the idempotent we can identify $\Lambda^2_D = (1-\Psi) T^2_D$. By this, one gains certain flexibility about bounded operators that span the set $\Lambda^2_D$. Of course, one can always choose $\Psi$ to generate the orthogonal complement, but we will see in Section \ref{VIB} that equating the algebraic and spectral torsion functionals is impossible in the Connes' second order differential calculus defined using the orthogonal complement.\\

There are two peculiarities about this approach from Mesland--Renie \cite{mesland2025existence}. One is that the idempotent $\Psi$ need not be unique. Secondly, it is harder to lift it to higher differential forms than the canonical Connes' construction and requires additional properties.




\section{The Finite geometry of the Standard model}

By the finite geometry of the Standard one usually refers to the spectral triple

\begin{equation}
    (A_\IR, \mathcal{H}, D)\;
\label{real SM}
\end{equation}
where $A_\IR$ is the the real algebra
\begin{equation}
    A_\IR = \mathbb{C}\oplus \mathbb{H}\oplus M_3(\mathbb{C}),
\end{equation}
the direct sum of the complex field, the field of quarternions and $M_3(\IC)$, which is represented on the 96 dimensional complex Hilbert space
\begin{equation}
    \mathcal{H} = \mathbb{C}^2\otimes \mathbb{C}^4\otimes\mathbb{C}^4\otimes\mathbb{C}^3
\label{Hilbert space}
\end{equation}
by the representation
\begin{equation}
    \begin{split}
	\pi(a) = \left[\begin{matrix}\lambda & 0\\0 & \overline{\lambda}\end{matrix}\right] \otimes \mathds{1}_4  \otimes e_{11}\otimes \mathds{1}_3 + \left[\begin{matrix}\alpha & \beta\\- \overline{\beta} & \overline{\alpha}\end{matrix}\right] \otimes \mathds{1}_4 \otimes e_{44}\otimes \mathds{1}_3 +\mathds{1}_2 \otimes \left[\begin{matrix}\lambda & 0 & 0 & 0\\0 & m_{11} & m_{12} & m_{13}\\0 & m_{21} & m_{22} & m_{23}\\0 & m_{31} & m_{32} & m_{33}\end{matrix}\right] \otimes \left(e_{22} + e_{33}\right)\otimes \mathds{1}_3 \;,\\
\end{split}
\label{pia real}
\end{equation}
and the Dirac operator
\begin{equation}
\begin{split}
    D &= e_{11}\otimes e_{11}\otimes \left[\begin{matrix}0 & 0 & \Upsilon_{R} & \Upsilon^{*}_{\nu}\\0 & 0 & \overline{\Upsilon}_\nu & 0\\\Upsilon^{*}_{R} & \Upsilon^{t}_{\nu} & 0 & 0\\\Upsilon_{\nu} & 0 & 0 & 0\end{matrix}\right]
 + e_{11}\otimes\left(\mathds{1}_4 - e_{11}\right)\otimes\left[\begin{matrix}0 & 0 & 0 & \Upsilon^{*}_{u}\\0 & 0 & \overline{\Upsilon}_u & 0\\0 & \Upsilon^{t}_{u} & 0 & 0\\\Upsilon_{u} & 0 & 0 & 0\end{matrix}\right]\\
 &+ e_{22}\otimes e_{11}\otimes \left[\begin{matrix}0 & 0 & 0 & \Upsilon^{*}_{e}\\0 & 0 & \overline{\Upsilon}_e & 0\\0 & \Upsilon^{t}_{e} & 0 & 0\\\Upsilon_{e} & 0 & 0 & 0\end{matrix}\right] + e_{22}\otimes\left(\mathds{1}_4 - e_{11}\right)\otimes\left[\begin{matrix}0 & 0 & 0 & \Upsilon^{*}_{d}\\0 & 0 & \overline{\Upsilon}_d & 0\\0 & \Upsilon^{t}_{d} & 0 & 0\\\Upsilon_{d} & 0 & 0 & 0\end{matrix}\right].
\end{split}
\label{Dirac operator}
\end{equation}

However, the theory of Connes and subsequent developments rely on the noncommutative algebras being modelled over the field of complex numbers. Hence we complexify \eqref{real SM} to the spectral triple
\begin{equation}
(A_F, \mathcal{H}_F, D_F)\;
\label{complexified SM}
\end{equation}
which amounts to complexifying the real algebra $A_\IR$ and its representation $\pi$ on the Hilbert space, while keeping the Dirac operator and Hilbert space the same, $D_F = D$, $\mathcal{H}_F = \mathcal{H}$. The process of complexification is performed summand--wise. It is well known that the complexification of the second direct summand $\mathbb{H}$ is $M_2(\mathbb{C})$, whereas the third summand $M_3(\mathbb{C})$ is already complexified to begin with. So we will only compute the complexification of the first summand, $\mathbb{C}$, and its representation. Given a real algebra $B$, the complexification $B_\IC$ is the range of the (surjective) multiplication map $\mu$
\[ \mu : B \otimes_\IR \IC \to \IC[B]. \]
Consider the representation of the real algebra $\IC$ (which appears in the first term of \eqref{pia real})
\begin{equation}
\pi(\lambda) = \left[\begin{matrix}\lambda & 0\\0 & \overline{\lambda}\end{matrix}\right]\;,
\label{real rep}
\end{equation}
then the range of the complexified representation of the complexified algebra $\IC_\IC$ is spanned by terms of the form
\begin{equation}
\mu \left(\left[\begin{matrix}\lambda & 0\\0 & \overline{\lambda}\end{matrix}\right]\otimes z\right) + \mu \left(\left[\begin{matrix}\mu & 0\\0 & \overline{\mu}\end{matrix}\right]\otimes w\right) = z\left[\begin{matrix}\lambda & 0\\0 & \overline{\lambda}\end{matrix}\right] + w\left[\begin{matrix}\mu & 0\\0 & \overline{\mu}\end{matrix}\right] \;,\quad \lambda,\mu,z,w\in\IC\;.
\label{complexification}
\end{equation}
It is easy to see that for 
$$z = \frac{\lambda \overline{\mu} - \mu^{2}}{\lambda \overline{\mu} - \mu \overline{\lambda}}\;,\quad w = \frac{\lambda \left(\mu - \overline{\lambda}\right)}{\lambda \overline{\mu} - \mu \overline{\lambda}}\;,
$$
the right-hand side of \eqref{complexification} becomes
$$\left[\begin{matrix}\lambda & 0\\0 & \mu\end{matrix}\right]$$
so it is clear that $\IC_\IC = \IC^2$ and we have determined the complexification of the representation \eqref{real rep}. All in all, the total complexified real algebra $A_F$ is\footnote{To understand why $\mathbb{C}_\IC = \mathbb{C}^2$, it is illuminating to consider a real basis of $\IC \otimes_\IR \IC$. It is given by $1\otimes_\IR 1, 1\otimes_\IR i, i\otimes_\IR 1$ and $i\otimes_\IR i$.}
\begin{equation}
    A_F = \mathbb{C}^2 \oplus M_2(\mathbb{C}) \oplus M_3(\mathbb{C})\;.
\label{AF algebra}
\end{equation}
with the complexified representation
\begin{equation}
    \begin{split}
	\pi^c(a) = \left[\begin{matrix}z & 0\\0 & w\end{matrix}\right] \otimes \mathds{1}_4  \otimes e_{11}\otimes \mathds{1}_3 + \left[\begin{matrix}\alpha & \beta\\ \gamma & \delta \end{matrix}\right] \otimes \mathds{1}_4 \otimes e_{44}\otimes \mathds{1}_3 +\mathds{1}_2 \otimes \left[\begin{matrix}z & 0 & 0 & 0\\0 & m_{11} & m_{12} & m_{13}\\0 & m_{21} & m_{22} & m_{23}\\0 & m_{31} & m_{32} & m_{33}\end{matrix}\right] \otimes \left(e_{22} + e_{33}\right)\otimes \mathds{1}_3 \\
\end{split}
\label{cpi_sm}
\end{equation}
Note that $z$ appears in two summands, while $w$ appears in only one summand in $\pi^c$. From here on, we will refer to \eqref{complexified SM} as the finite geometry of the SM.
\subsection{First order differential calculus}
Recalling the definition \eqref{d_D},\eqref{Omega1_D}, the first order differential calculus of the Standard Model is the data
\begin{equation}
    (\Omega^1_D(A_F), \dee_D)
\end{equation}
where $\dee_D(a) = [D,a]$ is the exterior derivative and $\Omega^1_D(A_F)$ is the bimodule over $A_F$ which is the linear space of elements $a[D,b]$. For an arbitrary element $a$
\begin{equation}
    \hspace{-10mm} a = \left[\begin{matrix}z\\w\end{matrix}\right]\oplus \left[\begin{matrix}\alpha & \beta\\\gamma & \delta\end{matrix}\right] \oplus \left[\begin{matrix}m_{11} & m_{12} & m_{13}\\m_{21} & m_{22} & m_{23}\\m_{31} & m_{32} & m_{33}\end{matrix}\right]
\label{a}
\end{equation}
of the complexified algebra $A_F$, its differential is given as the sum of four elements that are all mutually orthogonal (with respect to the standard scalar product in $B(\mathcal{H})$)
\begin{equation}
    \dee_D(a) = \left(\alpha - z\right)\omega_{\alpha - z} + \beta \omega_\beta + \gamma \omega_\gamma + \left(\delta - w\right)\omega_{\delta - w},
\label{da_psi}
\end{equation}
where the exact 1--forms appearing in \eqref{da_psi} read
\begin{equation}
\begin{array}{rr}
\begin{aligned}
\omega_{\alpha - z} = 
+&\left[\begin{matrix}1 & 0\\0 & 0\end{matrix}\right]\otimes\left[\begin{matrix}0 & 0 & 0 & 0\\0 & 1 & 0 & 0\\0 & 0 & 1 & 0\\0 & 0 & 0 & 1\end{matrix}\right]\otimes\left[\begin{matrix}0 & 0 & 0 & \Upsilon^{*}_{u}\\0 & 0 & 0 & 0\\0 & 0 & 0 & 0\\- \Upsilon_{u} & 0 & 0 & 0\end{matrix}\right]\\
+&\left[\begin{matrix}1 & 0\\0 & 0\end{matrix}\right]\otimes\left[\begin{matrix}1 & 0 & 0 & 0\\0 & 0 & 0 & 0\\0 & 0 & 0 & 0\\0 & 0 & 0 & 0\end{matrix}\right]\otimes\left[\begin{matrix}0 & 0 & 0 & \Upsilon^{*}_{\nu}\\0 & 0 & 0 & 0\\0 & 0 & 0 & 0\\- \Upsilon_{\nu} & 0 & 0 & 0\end{matrix}\right]\;,\\
\end{aligned}
&
\begin{aligned}
\omega_{\beta} = 
+&\left[\begin{matrix}0 & 1\\0 & 0\end{matrix}\right]\otimes\left[\begin{matrix}0 & 0 & 0 & 0\\0 & 1 & 0 & 0\\0 & 0 & 1 & 0\\0 & 0 & 0 & 1\end{matrix}\right]\otimes\left[\begin{matrix}0 & 0 & 0 & \Upsilon^{*}_{u}\\0 & 0 & 0 & 0\\0 & 0 & 0 & 0\\- \Upsilon_{d} & 0 & 0 & 0\end{matrix}\right]\\
+&\left[\begin{matrix}0 & 1\\0 & 0\end{matrix}\right]\otimes\left[\begin{matrix}1 & 0 & 0 & 0\\0 & 0 & 0 & 0\\0 & 0 & 0 & 0\\0 & 0 & 0 & 0\end{matrix}\right]\otimes\left[\begin{matrix}0 & 0 & 0 & \Upsilon^{*}_{\nu}\\0 & 0 & 0 & 0\\0 & 0 & 0 & 0\\- \Upsilon_{e} & 0 & 0 & 0\end{matrix}\right]\;,\\
\end{aligned}
\\
\begin{aligned}
\omega_{\gamma} = 
+&\left[\begin{matrix}0 & 0\\1 & 0\end{matrix}\right]\otimes\left[\begin{matrix}1 & 0 & 0 & 0\\0 & 0 & 0 & 0\\0 & 0 & 0 & 0\\0 & 0 & 0 & 0\end{matrix}\right]\otimes\left[\begin{matrix}0 & 0 & 0 & \Upsilon^{*}_{e}\\0 & 0 & 0 & 0\\0 & 0 & 0 & 0\\- \Upsilon_{\nu} & 0 & 0 & 0\end{matrix}\right]\\
+&\left[\begin{matrix}0 & 0\\1 & 0\end{matrix}\right]\otimes\left[\begin{matrix}0 & 0 & 0 & 0\\0 & 1 & 0 & 0\\0 & 0 & 1 & 0\\0 & 0 & 0 & 1\end{matrix}\right]\otimes\left[\begin{matrix}0 & 0 & 0 & \Upsilon^{*}_{d}\\0 & 0 & 0 & 0\\0 & 0 & 0 & 0\\- \Upsilon_{u} & 0 & 0 & 0\end{matrix}\right]\;,\\
\end{aligned}
&
\begin{aligned}
\omega_{\delta - w} = 
+&\left[\begin{matrix}0 & 0\\0 & 1\end{matrix}\right]\otimes\left[\begin{matrix}1 & 0 & 0 & 0\\0 & 0 & 0 & 0\\0 & 0 & 0 & 0\\0 & 0 & 0 & 0\end{matrix}\right]\otimes\left[\begin{matrix}0 & 0 & 0 & \Upsilon^{*}_{e}\\0 & 0 & 0 & 0\\0 & 0 & 0 & 0\\- \Upsilon_{e} & 0 & 0 & 0\end{matrix}\right]\\
+&\left[\begin{matrix}0 & 0\\0 & 1\end{matrix}\right]\otimes\left[\begin{matrix}0 & 0 & 0 & 0\\0 & 1 & 0 & 0\\0 & 0 & 1 & 0\\0 & 0 & 0 & 1\end{matrix}\right]\otimes\left[\begin{matrix}0 & 0 & 0 & \Upsilon^{*}_{d}\\0 & 0 & 0 & 0\\0 & 0 & 0 & 0\\- \Upsilon_{d} & 0 & 0 & 0\end{matrix}\right]\;.\\
\end{aligned}
\\
\end{array}
\end{equation}
As is evident from \eqref{da_psi}, $z$ and $\alpha$ always come together. The same also holds for $\delta$ and $w$, they come as $\delta - w$ in \eqref{da_psi}. Note the absence of the matrix $\Upsilon_R$. It makes sense that the differentials of represented algebra elements, which describe the Higgs mechanism of the full SM spectral triple\footnote{The full Standard Model spectral triple is defined as the spectral triple product of manifold's spectral triple and the internal spectral triple considered in this paper and defined by \eqref{complexified SM}.}, do not see the right-handed neutrino term $\Upsilon_R$ since Majorana mass term is a completely different way of generating mass to right handed neutrinos and it does not come from a Higgs mechanism. Similarly, the 1--forms of the internal part of the Standard Model's spectral triple \eqref{complexified SM} do not depend on $m\in M_3(\IC)$. This corresponds to the physical argument that the gluons are massless particles and as such they do not obtain masses through the Higgs mechanism.\\

We remark here that for $\alpha = z \neq 0,\ \beta = \gamma = \delta = w = 0, \ \dee_D(a)  =0$. An analogous result can be constructed for $\delta = w\neq 0$. One of the weakest notions of connectedness in NCG is the property that for the map $\dee_D : A \to \Omega^1_D$, $\ker(\dee_D) = \IC\cdot1_A$. Thus, we can conclude that our model is not connected in this weak sense.

Before proceeding further, we fix compressed notations for the exact 1--forms.\\

\begin{equation}
\begin{split}
\omega_{\gamma} = 
&+e_{2,1}\otimes  L \otimes \left(e_{1,4} \Upsilon^{*}_{e} - e_{4,1} \Upsilon_{\nu}\right)\\
&+e_{2,1}\otimes Q \otimes \left(e_{1,4} \Upsilon^{*}_{d} - e_{4,1} \Upsilon_{u}\right)\;,\\
\omega_{\delta - w} = 
&+e_{2,2}\otimes Q \otimes \left(e_{1,4} \Upsilon^{*}_{d} - e_{4,1} \Upsilon_{d}\right)\\
&+e_{2,2}\otimes  L \otimes \left(e_{1,4} \Upsilon^{*}_{e} - e_{4,1} \Upsilon_{e}\right)\;,\\
\omega_{\beta} = 
&+e_{1,2}\otimes Q \otimes \left(e_{1,4} \Upsilon^{*}_{u} - e_{4,1} \Upsilon_{d}\right)\\
&+e_{1,2}\otimes  L \otimes \left(e_{1,4} \Upsilon^{*}_{\nu} - e_{4,1} \Upsilon_{e}\right)\;,\\
\omega_{\alpha - z} = 
&+e_{1,1}\otimes Q \otimes \left(e_{1,4} \Upsilon^{*}_{u} - e_{4,1} \Upsilon_{u}\right)\\
&+e_{1,1}\otimes  L \otimes \left(e_{1,4} \Upsilon^{*}_{\nu} - e_{4,1} \Upsilon_{\nu}\right)\;.
\\
\label{exact 1--forms, compressed}
\end{split}
\end{equation}
Here the matrices $e_{i,j} = \delta_{ij}$ are to be understood as elements of $M_2(\IC)$ or $M_4(\IC)$ depending on the tensor leg they appear in.
We introduced symbols $Q = \mathds{1}_{4} - e_{1,1}$ and $L = e_{1,1}$ which appear only in the second legs in every summand of every 1--form in this paper. Moreover, the symbol $Q$ always appears in tensors whose last leg contains quark Yukawa couplings $\Upsilon_u,\Upsilon_d$ whereas $L$ appears in tensors with $\Upsilon_e$ or $\Upsilon_\nu$ Yukawa couplings in the last leg.\\
The space $\Omega^1_D$ is a free left $A_F$--module (as well as a free right module) generated by the four exact 1--forms, $\omega_{\alpha -z}, \omega_\beta, \omega_\gamma, \omega_{\delta -w}$. As a complex vector space, $\Omega^1_D$ is generated by the following linearly independent eight operators
\begin{equation}
\begin{array}{ll}
\vtop{\null\hbox{$\begin{aligned}
\omega_{1} = 
&+e_{1,1}\otimes  L \otimes e_{1,4} \Upsilon^{*}_{\nu}\\
&+e_{1,1}\otimes Q \otimes e_{1,4} \Upsilon^{*}_{u}\;,\\
\end{aligned}$}}
&
\vtop{\null\hbox{$\begin{aligned}
\omega_{2} = 
&+e_{1,2}\otimes Q \otimes e_{1,4} \Upsilon^{*}_{u}\\
&+e_{1,2}\otimes  L \otimes e_{1,4} \Upsilon^{*}_{\nu}\;,\\
\end{aligned}$}}
\\
\vtop{\null\hbox{$\begin{aligned}
\omega_{3} = 
&+e_{2,1}\otimes  L \otimes e_{1,4} \Upsilon^{*}_{e}\\
&+e_{2,1}\otimes Q \otimes e_{1,4} \Upsilon^{*}_{d}\;,\\
\end{aligned}$}}
&
\vtop{\null\hbox{$\begin{aligned}
\omega_{4} = 
&+e_{2,2}\otimes Q \otimes e_{1,4} \Upsilon^{*}_{d}\\
&+e_{2,2}\otimes  L \otimes e_{1,4} \Upsilon^{*}_{e}\;,\\
\end{aligned}$}}
\\
\vtop{\null\hbox{$\begin{aligned}
\omega_{5} = 
&+e_{1,1}\otimes Q \otimes e_{4,1} \Upsilon_{u}\\
&+e_{1,1}\otimes  L \otimes e_{4,1} \Upsilon_{\nu}\;,\\
\end{aligned}$}}
&
\vtop{\null\hbox{$\begin{aligned}
\omega_{6} = 
&+e_{2,1}\otimes  L \otimes e_{4,1} \Upsilon_{\nu}\\
&+e_{2,1}\otimes Q \otimes e_{4,1} \Upsilon_{u}\;,\\
\end{aligned}$}}
\\
\vtop{\null\hbox{$\begin{aligned}
\omega_{7} = 
&+e_{1,2}\otimes Q \otimes e_{4,1} \Upsilon_{d}\\
&+e_{1,2}\otimes  L \otimes e_{4,1} \Upsilon_{e}\;,\\
\end{aligned}$}}
&
\vtop{\null\hbox{$\begin{aligned}
\omega_{8} = 
&+e_{2,2}\otimes Q \otimes e_{4,1} \Upsilon_{d}\\
&+e_{2,2}\otimes  L \otimes e_{4,1} \Upsilon_{e}\;.\\
\end{aligned}$}}
\\
\end{array}
\label{basis over C 1--forms}
\end{equation}
Note that exact 1--forms $\omega$ in \eqref{exact 1--forms, compressed} split into the complex basis elements $\omega_i$ in \eqref{basis over C 1--forms} according to whether the last tensor leg is $e_{1,4}$ or $e_{4,1}$.
\section{Spectral functionals} \label{16sep25sm6}
Spectral quantities such as noncommutative integral and Ricci scalar curvature can be regarded as functionals
over $C^\infty(M)$ which often admit generalizations to noncommutative algebras. In \cite{Dabrowski:2022ufo}, and subsequently in \cite{DSZ24a} and
\cite{DSZ24b}, spectral functionals were constructed and studied which recover geometric data like metric, torsion and Einstein tensor from the Dirac operator of a spectral triple. In this section we consider also the spectral Ricci tensor functional.

In effect, we compute all these spectral functionals for the finite geometry of the Standard Model. Correspondingly in their definitions we set the manifold dimension $n=0$ and replace the usual Wodzicki residue by the (matrix) trace trace on $B(\mathcal{H})$ which is natural candidate for finite geometries in \cite[Chapter 6.3]{connes}. Then, in particular the noncommutative integral becomes on algebra elements $a\in A_F$
\begin{equation}
     \int\hspace{-3.7mm}-  a = \text{Tr}(a) .
\end{equation}
\subsection{Spectral metric functional}
\noindent
Let $u,v \in \Omega^1_D$ and let\footnote{When $D$ is clear from the context we will often omit $D$ as the subscript.} $\dee_D$ as in (5), such that
\begin{equation}
u = \sum_i a_i \dee(b_i)\;, \quad
v = \sum_j e_j \dee(f_j)\;,
\label{uv}
\end{equation}
where
\begin{equation}
\hspace{-10mm}
\begin{array}{lll}
&a_i = \left[\begin{matrix}z^{(i)}_1\\w^{(i)}_1\end{matrix}\right]\oplus \left[\begin{matrix}\alpha^{(i)}_1 & \beta^{(i)}_1\\\gamma^{(i)}_1 & \delta^{(i)}_1\end{matrix}\right] \oplus m^{(i)}_1\;, \quad  &b_i = \left[\begin{matrix}z^{(i)}_2\\w^{(i)}_2\end{matrix}\right]\oplus \left[\begin{matrix}\alpha^{(i)}_2 & \beta^{(i)}_2\\\gamma^{(i)}_2 & \delta^{(i)}_2\end{matrix}\right] \oplus m^{(i)}_2 \;,\\

&e_i = \left[\begin{matrix}z^{(j)}_3\\w^{(j)}_3\end{matrix}\right]\oplus \left[\begin{matrix}\alpha^{(j)}_3 & \beta^{(j)}_3\\\gamma^{(j)}_3 & \delta^{(j)}_3\end{matrix}\right] \oplus m^{(j)}_3\;,\quad &f_i = \left[\begin{matrix}z^{(j)}_4\\w^{(j)}_4\end{matrix}\right]\oplus \left[\begin{matrix}\alpha^{(j)}_4 & \beta^{(j)}_4\\\gamma^{(j)}_4 & \delta^{(j)}_4\end{matrix}\right] \oplus m^{(j)}_4 \;.\\
\end{array}
\label{abef}
\end{equation}
Then, the metric functional is given by
\begin{equation}
g_D(u,v) = \text{Tr}(uv)\;.
\end{equation}
Since the product of two 1--forms $u,v$ is an element of the bimodule $\pi_D(\Omega^2_u)$, it can always be expressed in the form $\sum_j a^{(j)} \dee(a^{(j)}_2) \dee(a^{(j)}_3)$. Thus the metric functional $g$ can be completely described by the following equation
\begin{equation}
\begin{split}
g(a\;\dee(a_2), \dee(a_3)) =& - \beta_{3} \left(\delta \gamma_{2} + \gamma \left(\alpha_{2} - z_{2}\right)\right) +\left[\Tr{\left(\Upsilon^{*}_{\nu} \Upsilon_{\nu} \right)}+3\Tr{\left(\Upsilon^{*}_{u} \Upsilon_{u} \right)}\right] \\
&- \gamma_{3} \left(\alpha \beta_{2} + \beta \left(\delta_{2} - w_{2}\right)\right)\left[\Tr{\left(\Upsilon^{*}_{e} \Upsilon_{e} \right)} +3\Tr{\left(\Upsilon^{*}_{d} \Upsilon_{d} \right)}\right]\\
&- w \left(\beta_{3} \gamma_{2} + \left(\delta_{2} - w_{2}\right) \left(\delta_{3} - w_{3}\right)\right)\left[\Tr{\left(\Upsilon^{*}_{e} \Upsilon_{e} \right)} +3\Tr{\left(\Upsilon^{*}_{d} \Upsilon_{d} \right)} \right]\\
&- z \left(\beta_{2} \gamma_{3} + \left(\alpha_{2} - z_{2}\right) \left(\alpha_{3} - z_{3}\right)\right)\left[ \Tr{\left(\Upsilon^{*}_{\nu} \Upsilon_{\nu} \right)}+3\Tr{\left(\Upsilon^{*}_{u} \Upsilon_{u} \right)}\right]\\
&- \left(\alpha_{3} - z_{3}\right) \left(\alpha \left(\alpha_{2} - z_{2}\right) + \beta \gamma_{2}\right)\left[\Tr{\left(\Upsilon^{*}_{\nu} \Upsilon_{\nu} \right)}+3\Tr{\left(\Upsilon^{*}_{u} \Upsilon_{u} \right)}\right]\\
&- \left(\delta_{3} - w_{3}\right) \left(\beta_{2} \gamma + \delta \left(\delta_{2} - w_{2}\right)\right)\left[ \Tr{\left(\Upsilon^{*}_{e} \Upsilon_{e} \right)}+3\Tr{\left(\Upsilon^{*}_{d} \Upsilon_{d} \right)}\right]\;,
\end{split}
\end{equation}
where the parameters without an index define the represented algebra element $a$, while parameters with indices $2$ and $3$ define $a_2$ and $a_3$ respectively.

\subsection{Spectral torsion functional}
\noindent
The specification of the spectral torsion functional in \cite{Dabrowski:2023irb} to finite geometries becomes a trilinear functional on 1--forms which reads $\Tr(uvwD)$ for $u, v, w\in \Omega^1_D(A)$. Similarly to the argument in the metric tensor functional, since the product of three 1--forms $u,v$ and $w$ is an element of the bimodule $\pi_D(\Omega^3_u)$, it can always be expressed in the form $\sum a^{(j)}_0 \dee (a^{(j)}_1) \dee (a^{(j)}_2) \dee (a^{(j)}_3)$. Thus the spectral torsion functional can be completely described by the following equation
\begin{equation} \label{16sep25sm7}
\begin{split}
\hspace{-10mm}
\text{Tr}( a_0 d(a_1) d(a_2) d(a_3) D) =&- \beta_{4} w \left(\gamma_{2} \left(\alpha_{3} - z_{3}\right) + \gamma_{3} \left(\delta_{2} - w_{2}\right)\right)\left[ \Tr{\left(\Upsilon^{*}_{\nu} \Upsilon_{e} \Upsilon^{*}_{e} \Upsilon_{\nu} \right)} + 3\Tr{\left(\Upsilon^{*}_{d} \Upsilon_{u} \Upsilon^{*}_{u} \Upsilon_{d} \right)}\right] \\
&- \gamma_{4} z \left(\beta_{2} \left(\delta_{3} - w_{3}\right) + \beta_{3} \left(\alpha_{2} - z_{2}\right)\right)\left[\Tr{\left(\Upsilon^{*}_{\nu} \Upsilon_{e} \Upsilon^{*}_{e} \Upsilon_{\nu} \right)}  +3\Tr{\left(\Upsilon^{*}_{d} \Upsilon_{u} \Upsilon^{*}_{u} \Upsilon_{d} \right)}\right]\\
&- w \left(\delta_{4} - w_{4}\right) \left(\beta_{3} \gamma_{2} + \left(\delta_{2} - w_{2}\right) \left(\delta_{3} - w_{3}\right)\right)\left[\Tr{\left(\left(\Upsilon^{*}_{e} \Upsilon_{e}\right)^{2} \right)} +3\Tr{\left(\left(\Upsilon^{*}_{d} \Upsilon_{d}\right)^{2} \right)} \right] \\
&- z \left(\alpha_{4} - z_{4}\right) \left(\beta_{2} \gamma_{3} + \left(\alpha_{2} - z_{2}\right) \left(\alpha_{3} - z_{3}\right)\right)\left[ \Tr{\left(\left(\Upsilon^{*}_{\nu} \Upsilon_{\nu}\right)^{2} \right)} +3\Tr{\left(\left(\Upsilon^{*}_{u} \Upsilon_{u}\right)^{2} \right)}\right] \\
&+ \left(\alpha \beta_{2} + \beta \left(\delta_{2} - w_{2}\right)\right) \left(\gamma_{3} \left(\alpha_{4} - z_{4}\right) + \gamma_{4} \left(\delta_{3} - w_{3}\right)\right)\left[\Tr{\left(\Upsilon^{*}_{\nu} \Upsilon_{e} \Upsilon^{*}_{e} \Upsilon_{\nu} \right)} +3\Tr{\left(\Upsilon^{*}_{d} \Upsilon_{u} \Upsilon^{*}_{u} \Upsilon_{d} \right)}\right]\\
&+ \left(\alpha \left(\alpha_{2} - z_{2}\right) + \beta \gamma_{2}\right) \left(\beta_{3} \gamma_{4} + \left(\alpha_{3} - z_{3}\right) \left(\alpha_{4} - z_{4}\right)\right) \left[\Tr{\left(\left(\Upsilon^{*}_{\nu} \Upsilon_{\nu}\right)^{2} \right)} +3\Tr{\left(\left(\Upsilon^{*}_{u} \Upsilon_{u}\right)^{2} \right)}\right]\\
&+  \left(\beta_{2} \gamma + \delta \left(\delta_{2} - w_{2}\right)\right) \left(\beta_{4} \gamma_{3} + \left(\delta_{3} - w_{3}\right) \left(\delta_{4} - w_{4}\right)\right)\left[\Tr{\left(\left(\Upsilon^{*}_{e} \Upsilon_{e}\right)^{2} \right)} +3\Tr{\left(\left(\Upsilon^{*}_{d} \Upsilon_{d}\right)^{2} \right)}\right]\\
&+ \left(\beta_{3} \left(\delta_{4} - w_{4}\right) + \beta_{4} \left(\alpha_{3} - z_{3}\right)\right) \left(\delta \gamma_{2} + \gamma \left(\alpha_{2} - z_{2}\right)\right)\left[ \Tr{\left(\Upsilon^{*}_{\nu} \Upsilon_{e} \Upsilon^{*}_{e} \Upsilon_{\nu} \right)}+3\Tr{\left(\Upsilon^{*}_{d} \Upsilon_{u} \Upsilon^{*}_{u} \Upsilon_{d} \right)}\right]\;.
\end{split}
\end{equation}
It is worth noting that the torsion functional $\Tr (uvw D)$ vanishes if and only if $uvw=0$, otherwise it is nonzero (for nonvanishing $\Upsilon$ parameters). The fact that \eqref{16sep25sm7} is nonzero is the first main result of this paper. Later we will prove that this torsional functional can be exactly recreated using the algebraic functionals in an algebraic second order differential calculus.
\subsection{Spectral Ricci scalar curvature functional}
Given the canonical (torsion free) Dirac operator $D_0$ on a spin manifold $M$, building on A. Connes' annoucements and \cite{Kalau:1993uc, Kastler:1993zj}, one defines a spectral Ricci scalar curvature functional $\mathcal{R}_{D_0}$ on $f\in C^\infty (M)$ from which the Ricci scalar curvature $R$ can be recovered.
For Dirac operator $D = D_0 - \frac{i}{8}\;^AT_{ijk}\gamma^i \gamma^j \gamma^k$
coupled to an antisymmetric torsion tensor $T_{ijk}$
a similar functional $\mathcal{R}_{D}$ was computed in \cite{Bochniak:2025rud}.
It agrees with the expression in \cite[Proposition 5.4]{PFAFFLE20121529} computed in terms of the second coefficient in heat kernel expansion (taking account of \cite{Ackermann:1994pg}). 
However, the relevant coefficients do not match the early computation in \cite{Kalau:1993uc}. (Note also these spectral formulae do not match the usual expression for the Ricci scalar curvature tensor computed in terms of a linear connection coupled to torsion, 
see e.g. \cite[Lemma 2.5]{PFAFFLE20121529}).
Since our spectral triple, importantly, corresponds to a finite noncommutative space with torsion, we are in fact computing $\mathcal{R}_{D}$ as in \cite{Bochniak:2025rud}, 
which reads in our case:
\begin{equation}
\begin{split}
\mathcal{R}_D(a) = \Tr(aD^2) = & \; \; 3 \left(\alpha +z\right)\Tr\left(\Upsilon_{u} \Upsilon^{*}_{u}\right) + 3 \left(\delta +w\right)\Tr\left(\Upsilon_{d} \Upsilon^{*}_{d}\right) + 2\Tr(m)\Tr\left(\Upsilon_{u} \Upsilon^{*}_u + \Upsilon_{d} \Upsilon^*_d\right) \\
&+\left(\alpha + 3z \right)\Tr(\Upsilon_{\nu} \Upsilon^{*}_{\nu})+\left(\delta + w + 2z\right) \Tr\left(\Upsilon_{e}\Upsilon^{*}_{e} \right) + 2z \Tr\left(\Upsilon_{R} \Upsilon^*_{R}\right)\;.
\end{split}
\label{R(a)}
\end{equation}

A few comments are in order here. As was already stated, $\mathcal{R}_D$ measures the Ricci scalar for spaces with torsion. From our torsion computation \eqref{16sep25sm7}, we know that the Standard Model Dirac operator $D$ \eqref{Dirac operator} produces vanishing torsion if and only if 
\begin{equation}
    \Upsilon_e = \Upsilon_\nu = \Upsilon_u = \Upsilon_d = 0_{3\times 3}\;,
\label{03x3}
\end{equation}
so the last term $2z \Tr(\Upsilon_R \Upsilon_R^*)$ in \eqref{R(a)}, 
which survives the condition \eqref{03x3}, definitely arises as a pure curvature (not torsion) contribution to $\mathcal{R}_{D}$. This immediately raises concerns about the possibility to equate the algebraic scalar curvature and spectral scalar curvature. 
Namely, as will be clear in Section \ref{VIB}, an algebraic scalar curvature defined along the lines\footnote{I.e., using tensor products of 1--forms, metric (algebra valued or scalar valued) and idempotent operators on tensor products of 1--forms.} of algebraic torsion, algebraic Riemann tensor and other tensors from \cite{mesland2025existence}, will not be possibly equal to  spectral Ricci scalar curvature functional due to the inherent blindness of first order differential calculus to the $\Upsilon_R$ term in the Dirac operator. In any case,  
the considered here spectral triple has nonzero Ricci scalar (of the corresponding connection coupled to torsion).\\

On a similar note, one can also notice that the matrix $m\in M_3(\IC)$, appearing in the general (complexified) algebra element \eqref{cpi_sm}, contributes to \eqref{R(a)} via its trace $\Tr(m)$
\begin{equation}
\mathcal{R}_D(a) \supset 2\Tr(m)\Tr\left(\Upsilon_{u} \Upsilon^{*}_u + \Upsilon_{d} \Upsilon^*_d\right)\;.
\end{equation}
This is the first time we encounter the matrix $m$ in the result of a computation in our paper. Similar to the $\Upsilon_R$ part of the Dirac operator which commutes with the entire represented algebra, the (represented) subalgebra $M_3(\mathbb{C})\leq A_F$ commutes with the whole Dirac operator, both with the $\Upsilon_R$ part and with all Yukawa couplings in $D$. This commuting of $M_3(\IC)$ with $D$ means that the first order differential calculus is blind to a part of the algebra and there are immediate difficulties in constructing algebraic functionals which would contain terms with $\Tr(m)$. \\

These raised points are difficult to resolve, but we believe that noncommutative geometry should definitely be improved in some way to address them. The reward could be important for physics. Namely, the problem mostly revolves around the Majorana mass matrix $\Upsilon_R$ whose presence in the Standard Model is so far under heavy theoretical and experimental debate. Additionally, should the internal Ricci curvature \eqref{R(a)} prove correct and phenomenologically accessible, it may even serve as a geometric measurement or constraint of the Majorana mass.

\subsection{Spectral Einstein and Ricci tensor functionals}
We come now to two quadratic spectral functionals which are susceptible to torsion.
The first is the Einstein tensor functional $\text{G}_D$ introduced in \cite{Dabrowski:2022ufo}, which  even in the case of a spin manifold $M$ depends on  the first jet of 1--forms, rather than on 1--forms themselves \cite{Bochniak:2024duf}. The other one, the Ricci tensor functional $\text{Ric}_D$, can be easily defined by modifying the coefficients of the two terms in the anticommutator present in $\text{G}_D$ and also classically depends on derivatives of 1-forms\footnote{See also \cite{Floricel:2016hnr} for a prior alternative approach.}. Nevertheless, simply for completion purposes, we compute them for finite geometries as follows.\\
 
The Einstein functional is given by
\begin{equation}
    G_D(u,v) = \Tr(u\{v,D\}D) = G'_D(u,v) + G''_D(u,v)\;,
\end{equation}
where
\begin{equation}
    G'_D(u,v) = \Tr(uvD^2)\;, \quad\quad G''_D(u,v) = \Tr(uDvD)\;,
\end{equation}
which we compute separately. Additionally, the $0$ dimensional analogue of Ricci tensor can be computed as 
\begin{equation}
\text{Ric}_D(u,v)= G'_D(u,v)+2G''_D(u,v)\;.
\end{equation}
For $G'_D(u,v)$ we have
\begin{equation} \footnotesize
\begin{split}
& G'_D(u,v) =\\
&+z_{1} \left(- \beta_{2} \left(\alpha_{4} \gamma_{3} + \delta_{3} \gamma_{4} - \gamma_{3} z_{4}\right) - \left(\alpha_{2} - z_{2}\right) \left(\alpha_{3} \alpha_{4} - \alpha_{3} z_{4} + \beta_{3} \gamma_{4}\right)\right) \Tr\left(\Upsilon^{*}_{\nu} \Upsilon_{\nu} \Upsilon_{R} \Upsilon^{*}_{R} + \Upsilon^{*}_{\nu} \Upsilon_{\nu}\Upsilon^{*}_{\nu} \Upsilon_{\nu}\right)\\
&-z_{3} \left(\alpha_{4} - z_{4}\right) \left(\alpha_{1} \alpha_{2} - \alpha_{1} z_{2} + \beta_{1} \gamma_{2}\right) \Tr\left(\Upsilon^{*}_{\nu} \Upsilon_{\nu} \Upsilon^{*}_{\nu} \Upsilon_{\nu} \right) + \left(- \beta_{4} z_{3} \left(\alpha_{2} \gamma_{1} + \delta_{1} \gamma_{2} - \gamma_{1} z_{2}\right) - \gamma_{4} w_{3} \left(\alpha_{1} \beta_{2} + \beta_{1} \delta_{2} - \beta_{1} w_{2}\right)\right) \Tr \left( \Upsilon_{e} \Upsilon^{*}_{e} \Upsilon_{\nu} \Upsilon^{*}_{\nu} \right) \\
&+3 \left(- \beta_{4} z_{3} \left(\alpha_{2} \gamma_{1} + \delta_{1} \gamma_{2} - \gamma_{1} z_{2}\right) - \gamma_{4} w_{3} \left(\alpha_{1} \beta_{2} + \beta_{1} \delta_{2} - \beta_{1} w_{2}\right)\right) \Tr \left( \Upsilon_{u} \Upsilon^{*}_{u} \Upsilon_{d} \Upsilon^{*}_{d} \right) \\
&+3 \left(- \beta_{2} z_{1} \left(\alpha_{4} \gamma_{3} + \delta_{3} \gamma_{4} - \gamma_{3} z_{4}\right) - z_{1} \left(\alpha_{2} - z_{2}\right) \left(\alpha_{3} \alpha_{4} - \alpha_{3} z_{4} + \beta_{3} \gamma_{4}\right) - z_{3} \left(\alpha_{4} - z_{4}\right) \left(\alpha_{1} \alpha_{2} - \alpha_{1} z_{2} + \beta_{1} \gamma_{2}\right)\right) \Tr\left(\Upsilon^{*}_{u} \Upsilon_{u} \Upsilon^{*}_{u} \Upsilon_{u}\right) \\
&+3 \left(- \gamma_{2} w_{1} \left(\alpha_{3} \beta_{4} + \beta_{3} \delta_{4} - \beta_{3} w_{4}\right) - w_{1} \left(\delta_{2} - w_{2}\right) \left(\beta_{4} \gamma_{3} + \delta_{3} \delta_{4} - \delta_{3} w_{4}\right) - w_{3} \left(\delta_{4} - w_{4}\right) \left(\beta_{2} \gamma_{1} + \delta_{1} \delta_{2} - \delta_{1} w_{2}\right)\right) \Tr\left(\Upsilon^{*}_{d} \Upsilon_{d} \Upsilon^{*}_{d} \Upsilon_{d}\right)\\
&+\left(- \gamma_{2} w_{1} \left(\alpha_{3} \beta_{4} + \beta_{3} \delta_{4} - \beta_{3} w_{4}\right) - w_{1} \left(\delta_{2} - w_{2}\right) \left(\beta_{4} \gamma_{3} + \delta_{3} \delta_{4} - \delta_{3} w_{4}\right) - w_{3} \left(\delta_{4} - w_{4}\right) \left(\beta_{2} \gamma_{1} + \delta_{1} \delta_{2} - \delta_{1} w_{2}\right)\right) \Tr\left(\Upsilon^{*}_{e} \Upsilon_{e} \Upsilon^{*}_{e} \Upsilon_{e}\right)
\end{split}
\end{equation}
For the combination of nonzero symbols $\gamma_1,\alpha_2,z_3, \beta_4 \neq 0$ (appearing in the first summand in $G'$) and all other parameters $=0$, we find
\begin{equation}
    G'_D(u,v) = - \alpha_{2} \beta_{4} \gamma_{1} z_{3} \Tr \left(\Upsilon_{\nu} \Upsilon^{*}_{\nu} \Upsilon_{e} \Upsilon^{*}_{e}\right) - 3 \alpha_{2} \beta_{4} \gamma_{1} z_{3} \Tr \left(\Upsilon_{u} \Upsilon^{*}_{u} \Upsilon_{d} \Upsilon^{*}_{d}\right).
\end{equation}
So $G'_D$ is not a trivial function. For $G''_D(u,v)$ we have
\begin{equation}
\footnotesize
\begin{split}
&G''(u,v) =\\
&+3 \left(w_{1} w_{3} \left(\delta_{2} - w_{2}\right) \left(\delta_{4} - w_{4}\right) + \left(\beta_{2} \gamma_{1} + \delta_{1} \delta_{2} - \delta_{1} w_{2}\right) \left(\beta_{4} \gamma_{3} + \delta_{3} \delta_{4} - \delta_{3} w_{4}\right)\right) \Tr\left(\Upsilon^{*}_{d} \Upsilon_{d} \Upsilon^{*}_{d} \Upsilon_{d}\right)\\
&+\left(w_{1} w_{3} \left(\delta_{2} - w_{2}\right) \left(\delta_{4} - w_{4}\right) + \left(\beta_{2} \gamma_{1} + \delta_{1} \delta_{2} - \delta_{1} w_{2}\right) \left(\beta_{4} \gamma_{3} + \delta_{3} \delta_{4} - \delta_{3} w_{4}\right)\right) \Tr\left(\Upsilon^{*}_{e} \Upsilon_{e} \Upsilon^{*}_{e} \Upsilon_{e}\right)\\
&+\left(z_{1} z_{3} \left(\alpha_{2} - z_{2}\right) \left(\alpha_{4} - z_{4}\right) + \left(\alpha_{1} \alpha_{2} - \alpha_{1} z_{2} + \beta_{1} \gamma_{2}\right) \left(\alpha_{3} \alpha_{4} - \alpha_{3} z_{4} + \beta_{3} \gamma_{4}\right)\right) \Tr\left(\Upsilon^{*}_{\nu} \Upsilon_{\nu} \Upsilon^{*}_{\nu} \Upsilon_{\nu}\right) + \\
&+3 \left(z_{1} z_{3} \left(\alpha_{2} - z_{2}\right) \left(\alpha_{4} - z_{4}\right) + \left(\alpha_{1} \alpha_{2} - \alpha_{1} z_{2} + \beta_{1} \gamma_{2}\right) \left(\alpha_{3} \alpha_{4} - \alpha_{3} z_{4} + \beta_{3} \gamma_{4}\right)\right) \Tr\left(\Upsilon^{*}_{u} \Upsilon_{u} \Upsilon^{*}_{u} \Upsilon_{u}\right) \\
&+\left(\beta_{2} \gamma_{4} w_{3} z_{1} + \beta_{4} \gamma_{2} w_{1} z_{3} + \left(\alpha_{1} \beta_{2} + \beta_{1} \delta_{2} - \beta_{1} w_{2}\right) \left(\alpha_{4} \gamma_{3} + \delta_{3} \gamma_{4} - \gamma_{3} z_{4}\right) + \left(\alpha_{2} \gamma_{1} + \delta_{1} \gamma_{2} - \gamma_{1} z_{2}\right) \left(\alpha_{3} \beta_{4} + \beta_{3} \delta_{4} - \beta_{3} w_{4}\right)\right) \Tr\left( \Upsilon_{e} \Upsilon^{*}_{e} \Upsilon_{\nu} \Upsilon^{*}_{\nu} \right) \\
&+3 \left(\beta_{2} \gamma_{4} w_{3} z_{1} + \beta_{4} \gamma_{2} w_{1} z_{3} + \left(\alpha_{1} \beta_{2} + \beta_{1} \delta_{2} - \beta_{1} w_{2}\right) \left(\alpha_{4} \gamma_{3} + \delta_{3} \gamma_{4} - \gamma_{3} z_{4}\right) + \left(\alpha_{2} \gamma_{1} + \delta_{1} \gamma_{2} - \gamma_{1} z_{2}\right) \left(\alpha_{3} \beta_{4} + \beta_{3} \delta_{4} - \beta_{3} w_{4}\right)\right) \Tr \left( \Upsilon_{u} \Upsilon^{*}_{u} \Upsilon_{d} \Upsilon^{*}_{d} \right)
\end{split}
\end{equation}

To see that $G''_D$ is not trivial, we the combination of nonzero symbols $\beta_2, \gamma_4, w_3,z_1 \neq 0$ (appearing in the second summand) and all other parameters $=0$. We get
\begin{equation}
G''_D(u,v) = \beta_{2} \gamma_{4} w_{3} z_{1}  \Tr \left(\Upsilon_{e} \Upsilon^{*}_{e} \Upsilon_{\nu} \Upsilon^{*}_{\nu}\right) + 3 \beta_{2} \gamma_{4} w_{3} z_{1}  \Tr \left(\Upsilon_{d} \Upsilon^{*}_{d} \Upsilon_{u} \Upsilon^{*}_{u}\right).
\end{equation}
We note that in the most general form of $G'_D$, the term $\Upsilon_R$ again appears, even though the description of 1--forms does not involve the term. Unlike in the case of Ricci scalar functional, this time one can not make as well grounded conclusions about restrictions on improvements of NC geometry, since the starting functionals have weaker properties.
\section{Second order differential calculi} Recall that in Section \ref{11sep25sm1}, we described two formalisms to construct second order differential calculi. In this section, we give the explicit description of both applied to the finite  geometry of the Standard Model.

\subsection{The Connes formalism}

We recall from \eqref{11sep25sm2}, that the Connes space of two forms is given by $\Omega^2_D := \pi_D(\Omega^2_u)/J^2_D$. The bimodule $\pi_D(\Omega^2_u)$ is generated as a free left $A_F$ module (as well as a free right module) by the following eight elements which are of the form $\sum_j [D,a_j][D,b_j]$ for some algebra elements $a_j, b_j$
\begin{equation} \label{16sep25sm2}
\begin{array}{ll}
\vtop{\null\hbox{$\begin{aligned}
\lambda_{\beta_{1} \gamma_{2}} = 
&+e_{1,1}\otimes  L \otimes \left(- e_{1,1} \Upsilon^{*}_{\nu} \Upsilon_{\nu} - e_{4,4} \Upsilon_{e} \Upsilon^{*}_{e}\right)\\
&+e_{1,1}\otimes Q \otimes \left(- e_{1,1} \Upsilon^{*}_{u} \Upsilon_{u} - e_{4,4} \Upsilon_{d} \Upsilon^{*}_{d}\right)\;,\\
\end{aligned}$}}
&
\vtop{\null\hbox{$\begin{aligned}
\lambda_{- \beta_{1} \delta_{2} + \beta_{1} w_{2}} = 
&+e_{1,2}\otimes Q \otimes \left(e_{1,1} \Upsilon^{*}_{u} \Upsilon_{d} + e_{4,4} \Upsilon_{d} \Upsilon^{*}_{d}\right)\\
&+e_{1,2}\otimes  L \otimes \left(e_{1,1} \Upsilon^{*}_{\nu} \Upsilon_{e} + e_{4,4} \Upsilon_{e} \Upsilon^{*}_{e}\right)\;,\\
\end{aligned}$}}
\\
\vtop{\null\hbox{$\begin{aligned}
\lambda_{- \delta_{1} \gamma_{2} + \gamma_{2} w_{1}} = 
&+e_{2,1}\otimes  L \otimes \left(e_{1,1} \Upsilon^{*}_{e} \Upsilon_{\nu} + e_{4,4} \Upsilon_{e} \Upsilon^{*}_{e}\right)\\
&+e_{2,1}\otimes Q \otimes \left(e_{1,1} \Upsilon^{*}_{d} \Upsilon_{u} + e_{4,4} \Upsilon_{d} \Upsilon^{*}_{d}\right)\;,\\
\end{aligned}$}}
&
\vtop{\null\hbox{$\begin{aligned}
\lambda_{- \delta_{1} \delta_{2} + \delta_{1} w_{2} + \delta_{2} w_{1} - w_{1} w_{2}} = 
&+e_{2,2}\otimes Q \otimes \left(e_{1,1} \Upsilon^{*}_{d} \Upsilon_{d} + e_{4,4} \Upsilon_{d} \Upsilon^{*}_{d}\right)\\
&+e_{2,2}\otimes  L \otimes \left(e_{1,1} \Upsilon^{*}_{e} \Upsilon_{e} + e_{4,4} \Upsilon_{e} \Upsilon^{*}_{e}\right)\\
\end{aligned}$}}
\\
\vtop{\null\hbox{$\begin{aligned}
\lambda_{- \alpha_{1} \alpha_{2} + \alpha_{1} z_{2} + \alpha_{2} z_{1} - z_{1} z_{2}} = 
&+e_{1,1}\otimes Q \otimes \left(e_{1,1} \Upsilon^{*}_{u} \Upsilon_{u} + e_{4,4} \Upsilon_{u} \Upsilon^{*}_{u}\right)\\
&+e_{1,1}\otimes  L \otimes \left(e_{1,1} \Upsilon^{*}_{\nu} \Upsilon_{\nu} + e_{4,4} \Upsilon_{\nu} \Upsilon^{*}_{\nu}\right)\;,\\
\end{aligned}$}}
&
\vtop{\null\hbox{$\begin{aligned}
\lambda_{\beta_{2} \gamma_{1}} = 
&+e_{2,2}\otimes  L \otimes \left(- e_{1,1} \Upsilon^{*}_{e} \Upsilon_{e} - e_{4,4} \Upsilon_{\nu} \Upsilon^{*}_{\nu}\right)\\
&+e_{2,2}\otimes Q \otimes \left(- e_{1,1} \Upsilon^{*}_{d} \Upsilon_{d} - e_{4,4} \Upsilon_{u} \Upsilon^{*}_{u}\right)\;,\\
\end{aligned}$}}
\\
\vtop{\null\hbox{$\begin{aligned}
\lambda_{- \alpha_{1} \beta_{2} + \beta_{2} z_{1}} = 
&+e_{1,2}\otimes  L \otimes \left(e_{1,1} \Upsilon^{*}_{\nu} \Upsilon_{e} + e_{4,4} \Upsilon_{\nu} \Upsilon^{*}_{\nu}\right)\\
&+e_{1,2}\otimes Q \otimes \left(e_{1,1} \Upsilon^{*}_{u} \Upsilon_{d} + e_{4,4} \Upsilon_{u} \Upsilon^{*}_{u}\right)\;,\\
\end{aligned}$}}
&
\vtop{\null\hbox{$\begin{aligned}
\lambda_{- \alpha_{2} \gamma_{1} + \gamma_{1} z_{2}} = 
&+e_{2,1}\otimes  L \otimes \left(e_{1,1} \Upsilon^{*}_{e} \Upsilon_{\nu} + e_{4,4} \Upsilon_{\nu} \Upsilon^{*}_{\nu}\right)\\
&+e_{2,1}\otimes Q \otimes \left(e_{1,1} \Upsilon^{*}_{d} \Upsilon_{u} + e_{4,4} \Upsilon_{u} \Upsilon^{*}_{u}\right)\;.\\
\end{aligned}$}}
\\
\end{array}
\end{equation}

As a vector space, $\pi_D(\Omega^2_u)$ has the following twelve dimensional basis.
\begin{equation}
\label{pi_D Omega^2_u}
\begin{array}{ll}
\vtop{\null\hbox{$\begin{aligned}
\Omega_{1} = 
&+e_{1,1}\otimes  L \otimes e_{4,4} \Upsilon_{e} \Upsilon^{*}_{e}\\
&+e_{1,1}\otimes Q \otimes e_{4,4} \Upsilon_{d} \Upsilon^{*}_{d}\;,\\
\end{aligned}$}}
&
\vtop{\null\hbox{$\begin{aligned}
\Omega_{2} = 
&+e_{1,2}\otimes  L \otimes e_{4,4} \Upsilon_{e} \Upsilon^{*}_{e}\\
&+e_{1,2}\otimes Q \otimes e_{4,4} \Upsilon_{d} \Upsilon^{*}_{d}\;,\\
\end{aligned}$}}
\\
\vtop{\null\hbox{$\begin{aligned}
\Omega_{3} = 
&+e_{2,1}\otimes  L \otimes e_{4,4} \Upsilon_{e} \Upsilon^{*}_{e}\\
&+e_{2,1}\otimes Q \otimes e_{4,4} \Upsilon_{d} \Upsilon^{*}_{d}\;,\\
\end{aligned}$}}
&
\vtop{\null\hbox{$\begin{aligned}
\Omega_{4} = 
&+e_{2,2}\otimes Q \otimes e_{4,4} \Upsilon_{d} \Upsilon^{*}_{d}\\
&+e_{2,2}\otimes  L \otimes e_{4,4} \Upsilon_{e} \Upsilon^{*}_{e}\;,\\
\end{aligned}$}}
\\
\vtop{\null\hbox{$\begin{aligned}
\Omega_{5} = 
&+e_{1,1}\otimes Q \otimes e_{4,4} \Upsilon_{u} \Upsilon^{*}_{u}\\
&+e_{1,1}\otimes  L \otimes e_{4,4} \Upsilon_{\nu} \Upsilon^{*}_{\nu}\;,\\
\end{aligned}$}}
&
\vtop{\null\hbox{$\begin{aligned}
\Omega_{6} = 
&+e_{1,2}\otimes  L \otimes e_{4,4} \Upsilon_{\nu} \Upsilon^{*}_{\nu}\\
&+e_{1,2}\otimes Q \otimes e_{4,4} \Upsilon_{u} \Upsilon^{*}_{u}\;,\\
\end{aligned}$}}
\\
\vtop{\null\hbox{$\begin{aligned}
\Omega_{7} = 
&+e_{2,1}\otimes Q \otimes e_{4,4} \Upsilon_{u} \Upsilon^{*}_{u}\\
&+e_{2,1}\otimes  L \otimes e_{4,4} \Upsilon_{\nu} \Upsilon^{*}_{\nu}\;,\\
\end{aligned}$}}
&
\vtop{\null\hbox{$\begin{aligned}
\Omega_{8} = 
&+e_{2,2}\otimes Q \otimes e_{4,4} \Upsilon_{u} \Upsilon^{*}_{u}\\
&+e_{2,2}\otimes  L \otimes e_{4,4} \Upsilon_{\nu} \Upsilon^{*}_{\nu}\;,\\
\end{aligned}$}}
\\
\vtop{\null\hbox{$\begin{aligned}
\Omega_{9} = 
&+e_{1,2}\otimes Q \otimes e_{1,1} \Upsilon^{*}_{u} \Upsilon_{d}\\
&+e_{1,2}\otimes  L \otimes e_{1,1} \Upsilon^{*}_{\nu} \Upsilon_{e}\;,\\
\end{aligned}$}}
&
\vtop{\null\hbox{$\begin{aligned}
\Omega_{10} = 
&+e_{2,1}\otimes Q \otimes e_{1,1} \Upsilon^{*}_{d} \Upsilon_{u}\\
&+e_{2,1}\otimes  L \otimes e_{1,1} \Upsilon^{*}_{e} \Upsilon_{\nu}\;,\\
\end{aligned}$}}
\\
\vtop{\null\hbox{$\begin{aligned}
\Omega_{11} = 
&+e_{2,2}\otimes  L \otimes e_{1,1} \Upsilon^{*}_{e} \Upsilon_{e}\\
&+e_{2,2}\otimes Q \otimes e_{1,1} \Upsilon^{*}_{d} \Upsilon_{d}\;,\\
\end{aligned}$}}
&
\vtop{\null\hbox{$\begin{aligned}
\Omega_{12} = 
&+e_{1,1}\otimes Q \otimes e_{1,1} \Upsilon^{*}_{u} \Upsilon_{u}\\
&+e_{1,1}\otimes  L \otimes e_{1,1} \Upsilon^{*}_{\nu} \Upsilon_{\nu}\;.\\
\end{aligned}$}}
\\
\end{array}
\end{equation}
Note that, similar to the case of 1--forms, the vector space basis is obtained from the decomposition of the third leg of $\lambda$'s \eqref{16sep25sm2}, according to whether they are $e_{1,1}$ or $e_{4,4}$. Actually, there are four basis elements with the component $e_{1,1}$ and eight with the component $e_{4,4}$. Next, we list a vector space basis for $J^2_D$, where, from definition,
\begin{equation}
    J^2_D = \quad_\mathbb{C}\langle \sum_j \dee(a_j) \dee(b_j) : (a_j,b_j) \text{ s.t. } \begin{array}{l}
        -\alpha_{1} \beta_{2} - \beta_{1} \delta_{2} + \beta_{1} w_{2} = 0\;, \\
       - \beta_{2} \gamma_{1} - \delta_{1} \delta_{2} + \delta_{1} w_{2} = 0\;, \\
       \gamma_{2} w_{1} = 0\;, \\
       \delta_{2} w_{1} - w_{1} w_{2} = 0\;, \\
       - \alpha_{1} \alpha_{2} + \alpha_{1} z_{2} - \beta_{1} \gamma_{2} = 0 \;,\\
       - \alpha_{2} \gamma_{1} - \delta_{1} \gamma_{2} + \gamma_{1} z_{2} = 0\;, \\
       \alpha_{2} z_{1} - z_{1} z_{2} = 0 \;,\\
       \beta_{2} z_{1} = 0 \;.\\
    \end{array}
    \rangle
\label{junkdef}
\end{equation}
(the restrictions are equations which set the 1--forms $\sum_j a_j \dee(b_j)$ to zero) 
\begin{equation} \label{16sep25sm3}
\begin{array}{ll}
\vtop{\null\hbox{$\begin{aligned}
\iota_{\beta_{1} \gamma_{2}} = 
&+e_{1,1}\otimes  L \otimes e_{4,4} \left(\Upsilon_{\nu} \Upsilon^{*}_{\nu} - \Upsilon_{e} \Upsilon^{*}_{e}\right)\\
&+e_{1,1}\otimes Q \otimes e_{4,4} \left(- \Upsilon_{d} \Upsilon^{*}_{d} + \Upsilon_{u} \Upsilon^{*}_{u}\right)\;,\\
\end{aligned}$}}
&
\vtop{\null\hbox{$\begin{aligned}
\iota_{\alpha_{1} \beta_{2}} = 
&+e_{1,2}\otimes Q \otimes e_{4,4} \left(\Upsilon_{d} \Upsilon^{*}_{d} - \Upsilon_{u} \Upsilon^{*}_{u}\right)\\
&+e_{1,2}\otimes  L \otimes e_{4,4} \left(- \Upsilon_{\nu} \Upsilon^{*}_{\nu} + \Upsilon_{e} \Upsilon^{*}_{e}\right)\;,\\
\end{aligned}$}}
\\
\vtop{\null\hbox{$\begin{aligned}
\iota_{\delta_{1} \gamma_{2}} = 
&+e_{2,1}\otimes  L \otimes e_{4,4} \left(\Upsilon_{\nu} \Upsilon^{*}_{\nu} - \Upsilon_{e} \Upsilon^{*}_{e}\right)\\
&+e_{2,1}\otimes Q \otimes e_{4,4} \left(- \Upsilon_{d} \Upsilon^{*}_{d} + \Upsilon_{u} \Upsilon^{*}_{u}\right)\;,\\
\end{aligned}$}}
&
\vtop{\null\hbox{$\begin{aligned}
\iota_{\beta_{2} \gamma_{1}} = 
&+e_{2,2}\otimes  L \otimes e_{4,4} \left(- \Upsilon_{\nu} \Upsilon^{*}_{\nu} + \Upsilon_{e} \Upsilon^{*}_{e}\right)\\
&+e_{2,2}\otimes Q \otimes e_{4,4} \left(\Upsilon_{d} \Upsilon^{*}_{d} - \Upsilon_{u} \Upsilon^{*}_{u}\right)\;.\\
\end{aligned}$}}
\\
\end{array}
\end{equation}
We observe that every junk form $\iota$ has $e_{4,4}$ as the last tensor leg, and moreover, each $\iota$ is formed as the difference of two distinct $\Omega$'s. Thus, the space of two forms $\Omega^2_D = \pi_D(\Omega^2_u)/J^2_D$ is eight dimensional as a vector space. Recall from the discussion in Section \ref{11sep25sm1}, that as a bimodule $\Omega^2_D$ is a space of classes of operators. However, the space of operators $\pi_D(\Omega^2_u)$ inherits a hermitian structure from the ambient space $B(H_F)$. So we can obtain the following orthogonal decomposition,
\begin{equation}
    \pi_D(\Omega^2_u) = J^2_D \bigoplus (J^2_D)^\perp,
\end{equation}
where $(J^2_D)^\perp \cong \Omega^2_D$ as bimodules. This isomorphism allows us to choose representatives from the classes of operators, which generate $(J^2_D)^\perp$ as a free left (as well as right) module. A particular generating set is given by
\begin{equation}
\begin{array}{l}
\begin{aligned}
&\lambda_{- \alpha_{1} \alpha_{2} + \alpha_{1} z_{2} + \alpha_{2} z_{1} - \beta_{1} \gamma_{2} - z_{1} z_{2}}\\ = 
&+e_{1,1}\otimes  L \otimes e_{1,1} \Upsilon^{*}_{\nu} \Upsilon_{\nu}\\
&+\frac{3 \Tr{\left(\left(\Upsilon^{*}_{d} \Upsilon_{d}\right)^{2} \right)} + \Tr{\left(\left(\Upsilon^{*}_{e} \Upsilon_{e}\right)^{2} \right)} - \Tr{\left(\Upsilon^{*}_{\nu} \Upsilon_{e} \Upsilon^{*}_{e} \Upsilon_{\nu} \right)} - 3 \Tr{\left(\Upsilon^{*}_{d} \Upsilon_{u} \Upsilon^{*}_{u} \Upsilon_{d} \right)}}{\Tr\left(\left(\Upsilon^*_e \Upsilon_e  - \Upsilon_\nu \Upsilon_\nu^*\right)^2 + 3\left(\Upsilon_u\Upsilon_u^* - \Upsilon_d \Upsilon_d^*\right)^2  \right)}e_{1,1}\otimes  L \otimes e_{4,4} \Upsilon_{\nu} \Upsilon^{*}_{\nu}\\
&+\frac{\Tr{\left(\left(\Upsilon^{*}_{\nu} \Upsilon_{\nu}\right)^{2} \right)} + 3 \Tr{\left(\left(\Upsilon^{*}_{u} \Upsilon_{u}\right)^{2} \right)} - \Tr{\left(\Upsilon^{*}_{\nu} \Upsilon_{e} \Upsilon^{*}_{e} \Upsilon_{\nu} \right)} - 3 \Tr{\left(\Upsilon^{*}_{d} \Upsilon_{u} \Upsilon^{*}_{u} \Upsilon_{d} \right)}}{\Tr\left(\left(\Upsilon^*_e \Upsilon_e  - \Upsilon_\nu \Upsilon_\nu^*\right)^2 + 3\left(\Upsilon_u\Upsilon_u^* - \Upsilon_d \Upsilon_d^*\right)^2  \right)}e_{1,1}\otimes  L \otimes e_{4,4} \Upsilon_{e} \Upsilon^{*}_{e}\\
&+e_{1,1}\otimes Q \otimes e_{1,1} \Upsilon^{*}_{u} \Upsilon_{u}\\
&+\frac{3 \Tr{\left(\left(\Upsilon^{*}_{d} \Upsilon_{d}\right)^{2} \right)} + \Tr{\left(\left(\Upsilon^{*}_{e} \Upsilon_{e}\right)^{2} \right)} - \Tr{\left(\Upsilon^{*}_{\nu} \Upsilon_{e} \Upsilon^{*}_{e} \Upsilon_{\nu} \right)} - 3 \Tr{\left(\Upsilon^{*}_{d} \Upsilon_{u} \Upsilon^{*}_{u} \Upsilon_{d} \right)}}{\Tr\left(\left(\Upsilon^*_e \Upsilon_e  - \Upsilon_\nu \Upsilon_\nu^*\right)^2 + 3\left(\Upsilon_u\Upsilon_u^* - \Upsilon_d \Upsilon_d^*\right)^2  \right)}e_{1,1}\otimes Q \otimes e_{4,4} \Upsilon_{u} \Upsilon^{*}_{u}\\
&+\frac{\Tr{\left(\left(\Upsilon^{*}_{\nu} \Upsilon_{\nu}\right)^{2} \right)} + 3 \Tr{\left(\left(\Upsilon^{*}_{u} \Upsilon_{u}\right)^{2} \right)} - \Tr{\left(\Upsilon^{*}_{\nu} \Upsilon_{e} \Upsilon^{*}_{e} \Upsilon_{\nu} \right)} - 3 \Tr{\left(\Upsilon^{*}_{d} \Upsilon_{u} \Upsilon^{*}_{u} \Upsilon_{d} \right)}}{\Tr\left(\left(\Upsilon^*_e \Upsilon_e  - \Upsilon_\nu \Upsilon_\nu^*\right)^2 + 3\left(\Upsilon_u\Upsilon_u^* - \Upsilon_d \Upsilon_d^*\right)^2  \right)}e_{1,1}\otimes Q \otimes e_{4,4} \Upsilon_{d} \Upsilon^{*}_{d}\;,\\
\end{aligned}
\\
\begin{aligned}
&\lambda_{- \alpha_{1} \beta_{2} - \beta_{1} \delta_{2} + \beta_{1} w_{2} + \beta_{2} z_{1}}\\ = 
&+e_{1,2}\otimes  L \otimes e_{1,1} \Upsilon^{*}_{\nu} \Upsilon_{e}\\
&+\frac{\Tr{\left(\left(\Upsilon^{*}_{\nu} \Upsilon_{\nu}\right)^{2} \right)} + 3 \Tr{\left(\left(\Upsilon^{*}_{u} \Upsilon_{u}\right)^{2} \right)} - \Tr{\left(\Upsilon^{*}_{\nu} \Upsilon_{e} \Upsilon^{*}_{e} \Upsilon_{\nu} \right)} - 3 \Tr{\left(\Upsilon^{*}_{d} \Upsilon_{u} \Upsilon^{*}_{u} \Upsilon_{d} \right)}}{\Tr\left(\left(\Upsilon^*_e \Upsilon_e  - \Upsilon_\nu \Upsilon_\nu^*\right)^2 + 3\left(\Upsilon_u\Upsilon_u^* - \Upsilon_d \Upsilon_d^*\right)^2  \right)}e_{1,2}\otimes  L \otimes e_{4,4} \Upsilon_{e} \Upsilon^{*}_{e}\\
&+\frac{3 \Tr{\left(\left(\Upsilon^{*}_{d} \Upsilon_{d}\right)^{2} \right)} + \Tr{\left(\left(\Upsilon^{*}_{e} \Upsilon_{e}\right)^{2} \right)} - \Tr{\left(\Upsilon^{*}_{\nu} \Upsilon_{e} \Upsilon^{*}_{e} \Upsilon_{\nu} \right)} - 3 \Tr{\left(\Upsilon^{*}_{d} \Upsilon_{u} \Upsilon^{*}_{u} \Upsilon_{d} \right)}}{\Tr\left(\left(\Upsilon^*_e \Upsilon_e  - \Upsilon_\nu \Upsilon_\nu^*\right)^2 + 3\left(\Upsilon_u\Upsilon_u^* - \Upsilon_d \Upsilon_d^*\right)^2  \right)}e_{1,2}\otimes  L \otimes e_{4,4} \Upsilon_{\nu} \Upsilon^{*}_{\nu}\\
&+e_{1,2}\otimes Q \otimes e_{1,1} \Upsilon^{*}_{u} \Upsilon_{d}\\
&+\frac{\Tr{\left(\left(\Upsilon^{*}_{\nu} \Upsilon_{\nu}\right)^{2} \right)} + 3 \Tr{\left(\left(\Upsilon^{*}_{u} \Upsilon_{u}\right)^{2} \right)} - \Tr{\left(\Upsilon^{*}_{\nu} \Upsilon_{e} \Upsilon^{*}_{e} \Upsilon_{\nu} \right)} - 3 \Tr{\left(\Upsilon^{*}_{d} \Upsilon_{u} \Upsilon^{*}_{u} \Upsilon_{d} \right)}}{\Tr\left(\left(\Upsilon^*_e \Upsilon_e  - \Upsilon_\nu \Upsilon_\nu^*\right)^2 + 3\left(\Upsilon_u\Upsilon_u^* - \Upsilon_d \Upsilon_d^*\right)^2  \right)}e_{1,2}\otimes Q \otimes e_{4,4} \Upsilon_{d} \Upsilon^{*}_{d}\\
&+\frac{3 \Tr{\left(\left(\Upsilon^{*}_{d} \Upsilon_{d}\right)^{2} \right)} + \Tr{\left(\left(\Upsilon^{*}_{e} \Upsilon_{e}\right)^{2} \right)} - \Tr{\left(\Upsilon^{*}_{\nu} \Upsilon_{e} \Upsilon^{*}_{e} \Upsilon_{\nu} \right)} - 3 \Tr{\left(\Upsilon^{*}_{d} \Upsilon_{u} \Upsilon^{*}_{u} \Upsilon_{d} \right)}}{\Tr\left(\left(\Upsilon^*_e \Upsilon_e  - \Upsilon_\nu \Upsilon_\nu^*\right)^2 + 3\left(\Upsilon_u\Upsilon_u^* - \Upsilon_d \Upsilon_d^*\right)^2  \right)}e_{1,2}\otimes Q \otimes e_{4,4} \Upsilon_{u} \Upsilon^{*}_{u}\;,\\
\end{aligned}
\end{array}
\end{equation}
\begin{equation}
\begin{array}{l}
\begin{aligned}
&\lambda_{- \alpha_{2} \gamma_{1} - \delta_{1} \gamma_{2} + \gamma_{1} z_{2} + \gamma_{2} w_{1}}\\ = 
&+e_{2,1}\otimes  L \otimes e_{1,1} \Upsilon^{*}_{e} \Upsilon_{\nu}\\
&+\frac{3 \Tr{\left(\left(\Upsilon^{*}_{d} \Upsilon_{d}\right)^{2} \right)} + \Tr{\left(\left(\Upsilon^{*}_{e} \Upsilon_{e}\right)^{2} \right)} - \Tr{\left(\Upsilon^{*}_{\nu} \Upsilon_{e} \Upsilon^{*}_{e} \Upsilon_{\nu} \right)} - 3 \Tr{\left(\Upsilon^{*}_{d} \Upsilon_{u} \Upsilon^{*}_{u} \Upsilon_{d} \right)}}{\Tr\left(\left(\Upsilon^*_e \Upsilon_e  - \Upsilon_\nu \Upsilon_\nu^*\right)^2 + 3\left(\Upsilon_u\Upsilon_u^* - \Upsilon_d \Upsilon_d^*\right)^2  \right)}e_{2,1}\otimes  L \otimes e_{4,4} \Upsilon_{\nu} \Upsilon^{*}_{\nu}\\
&+\frac{\Tr{\left(\left(\Upsilon^{*}_{\nu} \Upsilon_{\nu}\right)^{2} \right)} + 3 \Tr{\left(\left(\Upsilon^{*}_{u} \Upsilon_{u}\right)^{2} \right)} - \Tr{\left(\Upsilon^{*}_{\nu} \Upsilon_{e} \Upsilon^{*}_{e} \Upsilon_{\nu} \right)} - 3 \Tr{\left(\Upsilon^{*}_{d} \Upsilon_{u} \Upsilon^{*}_{u} \Upsilon_{d} \right)}}{\Tr\left(\left(\Upsilon^*_e \Upsilon_e  - \Upsilon_\nu \Upsilon_\nu^*\right)^2 + 3\left(\Upsilon_u\Upsilon_u^* - \Upsilon_d \Upsilon_d^*\right)^2  \right)}e_{2,1}\otimes  L \otimes e_{4,4} \Upsilon_{e} \Upsilon^{*}_{e}\\
&+e_{2,1}\otimes Q \otimes e_{1,1} \Upsilon^{*}_{d} \Upsilon_{u}\\
&+\frac{3 \Tr{\left(\left(\Upsilon^{*}_{d} \Upsilon_{d}\right)^{2} \right)} + \Tr{\left(\left(\Upsilon^{*}_{e} \Upsilon_{e}\right)^{2} \right)} - \Tr{\left(\Upsilon^{*}_{\nu} \Upsilon_{e} \Upsilon^{*}_{e} \Upsilon_{\nu} \right)} - 3 \Tr{\left(\Upsilon^{*}_{d} \Upsilon_{u} \Upsilon^{*}_{u} \Upsilon_{d} \right)}}{\Tr\left(\left(\Upsilon^*_e \Upsilon_e  - \Upsilon_\nu \Upsilon_\nu^*\right)^2 + 3\left(\Upsilon_u\Upsilon_u^* - \Upsilon_d \Upsilon_d^*\right)^2  \right)}e_{2,1}\otimes Q \otimes e_{4,4} \Upsilon_{u} \Upsilon^{*}_{u}\\
&+\frac{\Tr{\left(\left(\Upsilon^{*}_{\nu} \Upsilon_{\nu}\right)^{2} \right)} + 3 \Tr{\left(\left(\Upsilon^{*}_{u} \Upsilon_{u}\right)^{2} \right)} - \Tr{\left(\Upsilon^{*}_{\nu} \Upsilon_{e} \Upsilon^{*}_{e} \Upsilon_{\nu} \right)} - 3 \Tr{\left(\Upsilon^{*}_{d} \Upsilon_{u} \Upsilon^{*}_{u} \Upsilon_{d} \right)}}{\Tr\left(\left(\Upsilon^*_e \Upsilon_e  - \Upsilon_\nu \Upsilon_\nu^*\right)^2 + 3\left(\Upsilon_u\Upsilon_u^* - \Upsilon_d \Upsilon_d^*\right)^2  \right)}e_{2,1}\otimes Q \otimes e_{4,4} \Upsilon_{d} \Upsilon^{*}_{d}\;,\\
\end{aligned}
\\
\begin{aligned}
&\lambda_{- \beta_{2} \gamma_{1} - \delta_{1} \delta_{2} + \delta_{1} w_{2} + \delta_{2} w_{1} - w_{1} w_{2}}\\ = 
&+e_{2,2}\otimes  L \otimes e_{1,1} \Upsilon^{*}_{e} \Upsilon_{e}\\
&+\frac{\Tr{\left(\left(\Upsilon^{*}_{\nu} \Upsilon_{\nu}\right)^{2} \right)} + 3 \Tr{\left(\left(\Upsilon^{*}_{u} \Upsilon_{u}\right)^{2} \right)} - \Tr{\left(\Upsilon^{*}_{\nu} \Upsilon_{e} \Upsilon^{*}_{e} \Upsilon_{\nu} \right)} - 3 \Tr{\left(\Upsilon^{*}_{d} \Upsilon_{u} \Upsilon^{*}_{u} \Upsilon_{d} \right)}}{\Tr\left(\left(\Upsilon^*_e \Upsilon_e  - \Upsilon_\nu \Upsilon_\nu^*\right)^2 + 3\left(\Upsilon_u\Upsilon_u^* - \Upsilon_d \Upsilon_d^*\right)^2  \right)}e_{2,2}\otimes  L \otimes e_{4,4} \Upsilon_{e} \Upsilon^{*}_{e}\\
&+\frac{3 \Tr{\left(\left(\Upsilon^{*}_{d} \Upsilon_{d}\right)^{2} \right)} + \Tr{\left(\left(\Upsilon^{*}_{e} \Upsilon_{e}\right)^{2} \right)} - \Tr{\left(\Upsilon^{*}_{\nu} \Upsilon_{e} \Upsilon^{*}_{e} \Upsilon_{\nu} \right)} - 3 \Tr{\left(\Upsilon^{*}_{d} \Upsilon_{u} \Upsilon^{*}_{u} \Upsilon_{d} \right)}}{\Tr\left(\left(\Upsilon^*_e \Upsilon_e  - \Upsilon_\nu \Upsilon_\nu^*\right)^2 + 3\left(\Upsilon_u\Upsilon_u^* - \Upsilon_d \Upsilon_d^*\right)^2  \right)}e_{2,2}\otimes  L \otimes e_{4,4} \Upsilon_{\nu} \Upsilon^{*}_{\nu}\\
&+e_{2,2}\otimes Q \otimes e_{1,1} \Upsilon^{*}_{d} \Upsilon_{d}\\
&+\frac{\Tr{\left(\left(\Upsilon^{*}_{\nu} \Upsilon_{\nu}\right)^{2} \right)} + 3 \Tr{\left(\left(\Upsilon^{*}_{u} \Upsilon_{u}\right)^{2} \right)} - \Tr{\left(\Upsilon^{*}_{\nu} \Upsilon_{e} \Upsilon^{*}_{e} \Upsilon_{\nu} \right)} - 3 \Tr{\left(\Upsilon^{*}_{d} \Upsilon_{u} \Upsilon^{*}_{u} \Upsilon_{d} \right)}}{\Tr\left(\left(\Upsilon^*_e \Upsilon_e  - \Upsilon_\nu \Upsilon_\nu^*\right)^2 + 3\left(\Upsilon_u\Upsilon_u^* - \Upsilon_d \Upsilon_d^*\right)^2  \right)}e_{2,2}\otimes Q \otimes e_{4,4} \Upsilon_{d} \Upsilon^{*}_{d}\\
&+\frac{3 \Tr{\left(\left(\Upsilon^{*}_{d} \Upsilon_{d}\right)^{2} \right)} + \Tr{\left(\left(\Upsilon^{*}_{e} \Upsilon_{e}\right)^{2} \right)} - \Tr{\left(\Upsilon^{*}_{\nu} \Upsilon_{e} \Upsilon^{*}_{e} \Upsilon_{\nu} \right)} - 3 \Tr{\left(\Upsilon^{*}_{d} \Upsilon_{u} \Upsilon^{*}_{u} \Upsilon_{d} \right)}}{\Tr\left(\left(\Upsilon^*_e \Upsilon_e  - \Upsilon_\nu \Upsilon_\nu^*\right)^2 + 3\left(\Upsilon_u\Upsilon_u^* - \Upsilon_d \Upsilon_d^*\right)^2  \right)}e_{2,2}\otimes Q \otimes e_{4,4} \Upsilon_{u} \Upsilon^{*}_{u}\;.\\
\end{aligned}
\end{array}
\end{equation}
From this set one can also read off the generating set over the complex numbers. Simply split every algebra generator $\lambda$ into two complex generators -- one with all terms with $e_{1,1}$ in the third leg, and one with all $e_{4,4}$ terms in the third leg. For example, the first generator over the algebra, $\lambda_{- \alpha_{1} \alpha_{2} + \alpha_{1} z_{2} + \alpha_{2} z_{1} - \beta_{1} \gamma_{2} - z_{1} z_{2}}$, splits into two generators over $\IC$, $\lambda_1$ and $\lambda_2$, given as:
\begin{equation}
\label{C twoforms basis}
\begin{array}{l}
\begin{aligned}
\lambda_1 = 
&+e_{1,1}\otimes  L \otimes e_{1,1} \Upsilon^{*}_{\nu} \Upsilon_{\nu}\\
&+e_{1,1}\otimes Q \otimes e_{1,1} \Upsilon^{*}_{u} \Upsilon_{u}\;,\\
\end{aligned}
\\
\begin{aligned}
\lambda_2 =
&+\frac{3 \Tr{\left(\left(\Upsilon^{*}_{d} \Upsilon_{d}\right)^{2} \right)} + \Tr{\left(\left(\Upsilon^{*}_{e} \Upsilon_{e}\right)^{2} \right)} - \Tr{\left(\Upsilon^{*}_{\nu} \Upsilon_{e} \Upsilon^{*}_{e} \Upsilon_{\nu} \right)} - 3 \Tr{\left(\Upsilon^{*}_{d} \Upsilon_{u} \Upsilon^{*}_{u} \Upsilon_{d} \right)}}{\Tr\left(\left(\Upsilon^*_e \Upsilon_e  - \Upsilon_\nu \Upsilon_\nu^*\right)^2 + 3\left(\Upsilon_u\Upsilon_u^* - \Upsilon_d \Upsilon_d^*\right)^2  \right)}e_{1,1}\otimes  L \otimes e_{4,4} \Upsilon_{\nu} \Upsilon^{*}_{\nu}\\
&+\frac{\Tr{\left(\left(\Upsilon^{*}_{\nu} \Upsilon_{\nu}\right)^{2} \right)} + 3 \Tr{\left(\left(\Upsilon^{*}_{u} \Upsilon_{u}\right)^{2} \right)} - \Tr{\left(\Upsilon^{*}_{\nu} \Upsilon_{e} \Upsilon^{*}_{e} \Upsilon_{\nu} \right)} - 3 \Tr{\left(\Upsilon^{*}_{d} \Upsilon_{u} \Upsilon^{*}_{u} \Upsilon_{d} \right)}}{\Tr\left(\left(\Upsilon^*_e \Upsilon_e  - \Upsilon_\nu \Upsilon_\nu^*\right)^2 + 3\left(\Upsilon_u\Upsilon_u^* - \Upsilon_d \Upsilon_d^*\right)^2  \right)}e_{1,1}\otimes  L \otimes e_{4,4} \Upsilon_{e} \Upsilon^{*}_{e}\\
&+\frac{3 \Tr{\left(\left(\Upsilon^{*}_{d} \Upsilon_{d}\right)^{2} \right)} + \Tr{\left(\left(\Upsilon^{*}_{e} \Upsilon_{e}\right)^{2} \right)} - \Tr{\left(\Upsilon^{*}_{\nu} \Upsilon_{e} \Upsilon^{*}_{e} \Upsilon_{\nu} \right)} - 3 \Tr{\left(\Upsilon^{*}_{d} \Upsilon_{u} \Upsilon^{*}_{u} \Upsilon_{d} \right)}}{\Tr\left(\left(\Upsilon^*_e \Upsilon_e  - \Upsilon_\nu \Upsilon_\nu^*\right)^2 + 3\left(\Upsilon_u\Upsilon_u^* - \Upsilon_d \Upsilon_d^*\right)^2  \right)}e_{1,1}\otimes Q \otimes e_{4,4} \Upsilon_{u} \Upsilon^{*}_{u}\\
&+\frac{\Tr{\left(\left(\Upsilon^{*}_{\nu} \Upsilon_{\nu}\right)^{2} \right)} + 3 \Tr{\left(\left(\Upsilon^{*}_{u} \Upsilon_{u}\right)^{2} \right)} - \Tr{\left(\Upsilon^{*}_{\nu} \Upsilon_{e} \Upsilon^{*}_{e} \Upsilon_{\nu} \right)} - 3 \Tr{\left(\Upsilon^{*}_{d} \Upsilon_{u} \Upsilon^{*}_{u} \Upsilon_{d} \right)}}{\Tr\left(\left(\Upsilon^*_e \Upsilon_e  - \Upsilon_\nu \Upsilon_\nu^*\right)^2 + 3\left(\Upsilon_u\Upsilon_u^* - \Upsilon_d \Upsilon_d^*\right)^2  \right)}e_{1,1}\otimes Q \otimes e_{4,4} \Upsilon_{d} \Upsilon^{*}_{d}\;,\\
\end{aligned}
\end{array}
\end{equation}
and similarly for all other elements of the generating set.
\subsection{The Mesland--Rennie formalism}\label{Sec V.B.}
We recall from \eqref{11sep25sm3}, that the Mesland--Rennie space of two forms is given by $\Lambda^2_D := T^2_D/JT^2_D$. From \eqref{16sep25sm2}, we know that the bimodule $\pi_D(\Omega^2_u)$ as a complex vectors space is twelve dimensional. By similar direct computations, the bimodule $T^2_D$ as a complex vector space is also twelve dimensional. However, from discussions in Section \ref{11sep25sm1}, the multiplication map $\mu: T^2_D \to \pi_D(\Omega^2_D)$ is surjective. Hence, by the rank nullity theorem for vector spaces, the map $\mu$ is a bijective map. Further more the restricted multiplication map $\mu : JT^2_D \to J^2_D$ is also surjective. Thus, $\mu|_{JT^2_D}$ is bijective onto $J^2_D$, and $JT^2_D = \mu^{-1}(J^2_D)$. From \eqref{16sep25sm3}, we have a complex vector space basis for $J^2_D$. After computing the pre--image of each basis element of $J^2_D$, we reorder to the following Hermitian basis for $JT^2_D = \mu^{-1}(J^2_D)$ with respect to the hermitian conjugation $^*$ given in \cite{mesland2025existence}
$$T^2_D \ni \left(\omega_1 \otimes \omega_2\right)^* = \omega_2^* \otimes \omega_1^* $$

\begin{equation}
\begin{array}{ll}
\vtop{\null\hbox{$\begin{aligned}
\mathfrak{h}_{1} = 
&-e_{1,1}\otimes  L \otimes e_{4,1}\Upsilon_{\nu}\otimes e_{1,1}\otimes  L \otimes e_{1,4}\Upsilon^{*}_{\nu}\\
&-e_{1,1}\otimes  L \otimes e_{4,1}\Upsilon_{\nu}\otimes e_{1,1}\otimes Q \otimes e_{1,4}\Upsilon^{*}_{u}\\
&-e_{1,1}\otimes Q \otimes e_{4,1}\Upsilon_{u}\otimes e_{1,1}\otimes  L \otimes e_{1,4}\Upsilon^{*}_{\nu}\\
&-e_{1,1}\otimes Q \otimes e_{4,1}\Upsilon_{u}\otimes e_{1,1}\otimes Q \otimes e_{1,4}\Upsilon^{*}_{u}\\
&+e_{1,2}\otimes  L \otimes e_{4,1}\Upsilon_{e}\otimes e_{2,1}\otimes  L \otimes e_{1,4}\Upsilon^{*}_{e}\\
&+e_{1,2}\otimes  L \otimes e_{4,1}\Upsilon_{e}\otimes e_{2,1}\otimes Q \otimes e_{1,4}\Upsilon^{*}_{d}\\
&+e_{1,2}\otimes Q \otimes e_{4,1}\Upsilon_{d}\otimes e_{2,1}\otimes  L \otimes e_{1,4}\Upsilon^{*}_{e}\\
&+e_{1,2}\otimes Q \otimes e_{4,1}\Upsilon_{d}\otimes e_{2,1}\otimes Q \otimes e_{1,4}\Upsilon^{*}_{d}\;,\\
\end{aligned}$}}
&
\vtop{\null\hbox{$\begin{aligned}
\mathfrak{h}_{2} = 
&-e_{1,1}\otimes  L \otimes e_{4,1}\Upsilon_{\nu}\otimes e_{1,2}\otimes  L \otimes e_{1,4}\Upsilon^{*}_{\nu}\\
&-e_{1,1}\otimes  L \otimes e_{4,1}\Upsilon_{\nu}\otimes e_{1,2}\otimes Q \otimes e_{1,4}\Upsilon^{*}_{u}\\
&-e_{1,1}\otimes Q \otimes e_{4,1}\Upsilon_{u}\otimes e_{1,2}\otimes  L \otimes e_{1,4}\Upsilon^{*}_{\nu}\\
&-e_{1,1}\otimes Q \otimes e_{4,1}\Upsilon_{u}\otimes e_{1,2}\otimes Q \otimes e_{1,4}\Upsilon^{*}_{u}\\
&+e_{1,2}\otimes  L \otimes e_{4,1}\Upsilon_{e}\otimes e_{2,2}\otimes  L \otimes e_{1,4}\Upsilon^{*}_{e}\\
&+e_{1,2}\otimes  L \otimes e_{4,1}\Upsilon_{e}\otimes e_{2,2}\otimes Q \otimes e_{1,4}\Upsilon^{*}_{d}\\
&+e_{1,2}\otimes Q \otimes e_{4,1}\Upsilon_{d}\otimes e_{2,2}\otimes  L \otimes e_{1,4}\Upsilon^{*}_{e}\\
&+e_{1,2}\otimes Q \otimes e_{4,1}\Upsilon_{d}\otimes e_{2,2}\otimes Q \otimes e_{1,4}\Upsilon^{*}_{d}\\
&-e_{2,1}\otimes  L \otimes e_{4,1}\Upsilon_{\nu}\otimes e_{1,1}\otimes  L \otimes e_{1,4}\Upsilon^{*}_{\nu}\\
&-e_{2,1}\otimes  L \otimes e_{4,1}\Upsilon_{\nu}\otimes e_{1,1}\otimes Q \otimes e_{1,4}\Upsilon^{*}_{u}\\
&-e_{2,1}\otimes Q \otimes e_{4,1}\Upsilon_{u}\otimes e_{1,1}\otimes  L \otimes e_{1,4}\Upsilon^{*}_{\nu}\\
&-e_{2,1}\otimes Q \otimes e_{4,1}\Upsilon_{u}\otimes e_{1,1}\otimes Q \otimes e_{1,4}\Upsilon^{*}_{u}\\
&+e_{2,2}\otimes  L \otimes e_{4,1}\Upsilon_{e}\otimes e_{2,1}\otimes  L \otimes e_{1,4}\Upsilon^{*}_{e}\\
&+e_{2,2}\otimes  L \otimes e_{4,1}\Upsilon_{e}\otimes e_{2,1}\otimes Q \otimes e_{1,4}\Upsilon^{*}_{d}\\
&+e_{2,2}\otimes Q \otimes e_{4,1}\Upsilon_{d}\otimes e_{2,1}\otimes  L \otimes e_{1,4}\Upsilon^{*}_{e}\\
&+e_{2,2}\otimes Q \otimes e_{4,1}\Upsilon_{d}\otimes e_{2,1}\otimes Q \otimes e_{1,4}\Upsilon^{*}_{d}\;,\\
\end{aligned}$}}
\\
\vtop{\null\hbox{$\begin{aligned}
\mathfrak{h}_{3} = 
&- ie_{1,1}\otimes  L \otimes e_{4,1}\Upsilon_{\nu}\otimes e_{1,2}\otimes  L \otimes e_{1,4}\Upsilon^{*}_{\nu}\\
&- ie_{1,1}\otimes  L \otimes e_{4,1}\Upsilon_{\nu}\otimes e_{1,2}\otimes Q \otimes e_{1,4}\Upsilon^{*}_{u}\\
&- ie_{1,1}\otimes Q \otimes e_{4,1}\Upsilon_{u}\otimes e_{1,2}\otimes  L \otimes e_{1,4}\Upsilon^{*}_{\nu}\\
&- ie_{1,1}\otimes Q \otimes e_{4,1}\Upsilon_{u}\otimes e_{1,2}\otimes Q \otimes e_{1,4}\Upsilon^{*}_{u}\\
&+ie_{1,2}\otimes  L \otimes e_{4,1}\Upsilon_{e}\otimes e_{2,2}\otimes  L \otimes e_{1,4}\Upsilon^{*}_{e}\\
&+ie_{1,2}\otimes  L \otimes e_{4,1}\Upsilon_{e}\otimes e_{2,2}\otimes Q \otimes e_{1,4}\Upsilon^{*}_{d}\\
&+ie_{1,2}\otimes Q \otimes e_{4,1}\Upsilon_{d}\otimes e_{2,2}\otimes  L \otimes e_{1,4}\Upsilon^{*}_{e}\\
&+ie_{1,2}\otimes Q \otimes e_{4,1}\Upsilon_{d}\otimes e_{2,2}\otimes Q \otimes e_{1,4}\Upsilon^{*}_{d}\\
&+ie_{2,1}\otimes  L \otimes e_{4,1}\Upsilon_{\nu}\otimes e_{1,1}\otimes  L \otimes e_{1,4}\Upsilon^{*}_{\nu}\\
&+ie_{2,1}\otimes  L \otimes e_{4,1}\Upsilon_{\nu}\otimes e_{1,1}\otimes Q \otimes e_{1,4}\Upsilon^{*}_{u}\\
&+ie_{2,1}\otimes Q \otimes e_{4,1}\Upsilon_{u}\otimes e_{1,1}\otimes  L \otimes e_{1,4}\Upsilon^{*}_{\nu}\\
&+ie_{2,1}\otimes Q \otimes e_{4,1}\Upsilon_{u}\otimes e_{1,1}\otimes Q \otimes e_{1,4}\Upsilon^{*}_{u}\\
&- ie_{2,2}\otimes  L \otimes e_{4,1}\Upsilon_{e}\otimes e_{2,1}\otimes  L \otimes e_{1,4}\Upsilon^{*}_{e}\\
&- ie_{2,2}\otimes  L \otimes e_{4,1}\Upsilon_{e}\otimes e_{2,1}\otimes Q \otimes e_{1,4}\Upsilon^{*}_{d}\\
&- ie_{2,2}\otimes Q \otimes e_{4,1}\Upsilon_{d}\otimes e_{2,1}\otimes  L \otimes e_{1,4}\Upsilon^{*}_{e}\\
&- ie_{2,2}\otimes Q \otimes e_{4,1}\Upsilon_{d}\otimes e_{2,1}\otimes Q \otimes e_{1,4}\Upsilon^{*}_{d}\;,\\
\end{aligned}$}}
&
\vtop{\null\hbox{$\begin{aligned}
\mathfrak{h}_{4} = 
&-e_{2,1}\otimes  L \otimes e_{4,1}\Upsilon_{\nu}\otimes e_{1,2}\otimes  L \otimes e_{1,4}\Upsilon^{*}_{\nu}\\
&-e_{2,1}\otimes  L \otimes e_{4,1}\Upsilon_{\nu}\otimes e_{1,2}\otimes Q \otimes e_{1,4}\Upsilon^{*}_{u}\\
&-e_{2,1}\otimes Q \otimes e_{4,1}\Upsilon_{u}\otimes e_{1,2}\otimes  L \otimes e_{1,4}\Upsilon^{*}_{\nu}\\
&-e_{2,1}\otimes Q \otimes e_{4,1}\Upsilon_{u}\otimes e_{1,2}\otimes Q \otimes e_{1,4}\Upsilon^{*}_{u}\\
&+e_{2,2}\otimes  L \otimes e_{4,1}\Upsilon_{e}\otimes e_{2,2}\otimes  L \otimes e_{1,4}\Upsilon^{*}_{e}\\
&+e_{2,2}\otimes  L \otimes e_{4,1}\Upsilon_{e}\otimes e_{2,2}\otimes Q \otimes e_{1,4}\Upsilon^{*}_{d}\\
&+e_{2,2}\otimes Q \otimes e_{4,1}\Upsilon_{d}\otimes e_{2,2}\otimes  L \otimes e_{1,4}\Upsilon^{*}_{e}\\
&+e_{2,2}\otimes Q \otimes e_{4,1}\Upsilon_{d}\otimes e_{2,2}\otimes Q \otimes e_{1,4}\Upsilon^{*}_{d}\;.\\
\end{aligned}$}}
\\
\end{array}
\end{equation}
The orthogonal complement of $JT^2_D$ in $T^2_D$ is thus eight dimensional. The following eight Hermitian elements, together with the four $\frak{h}$ together span the entirety of $T^2_D$
\begin{equation}
\begin{array}{ll}
\vtop{\null\hbox{$\begin{aligned}
h_{1} = 
&+e_{1,1}\otimes  L \otimes e_{4,1} \Upsilon_{\nu}\otimes e_{1,1}\otimes  L \otimes e_{1,4} \Upsilon^{*}_{\nu}\\
&+e_{1,1}\otimes  L \otimes e_{4,1} \Upsilon_{\nu}\otimes e_{1,1}\otimes Q \otimes e_{1,4} \Upsilon^{*}_{u}\\
&+e_{1,1}\otimes Q \otimes e_{4,1} \Upsilon_{u}\otimes e_{1,1}\otimes  L \otimes e_{1,4} \Upsilon^{*}_{\nu}\\
&+e_{1,1}\otimes Q \otimes e_{4,1} \Upsilon_{u}\otimes e_{1,1}\otimes Q \otimes e_{1,4} \Upsilon^{*}_{u}\\
&+e_{1,2}\otimes  L \otimes e_{4,1} \Upsilon_{e}\otimes e_{2,1}\otimes  L \otimes e_{1,4} \Upsilon^{*}_{e}\\
&+e_{1,2}\otimes  L \otimes e_{4,1} \Upsilon_{e}\otimes e_{2,1}\otimes Q \otimes e_{1,4} \Upsilon^{*}_{d}\\
&+e_{1,2}\otimes Q \otimes e_{4,1} \Upsilon_{d}\otimes e_{2,1}\otimes  L \otimes e_{1,4} \Upsilon^{*}_{e}\\
&+e_{1,2}\otimes Q \otimes e_{4,1} \Upsilon_{d}\otimes e_{2,1}\otimes Q \otimes e_{1,4} \Upsilon^{*}_{d}\;,\\
\end{aligned}$}}
&
\vtop{\null\hbox{$\begin{aligned}
h_{2} = 
&+e_{1,1}\otimes  L \otimes e_{4,1} \Upsilon_{\nu}\otimes e_{1,2}\otimes  L \otimes e_{1,4} \Upsilon^{*}_{\nu}\\
&+e_{1,1}\otimes  L \otimes e_{4,1} \Upsilon_{\nu}\otimes e_{1,2}\otimes Q \otimes e_{1,4} \Upsilon^{*}_{u}\\
&+e_{1,1}\otimes Q \otimes e_{4,1} \Upsilon_{u}\otimes e_{1,2}\otimes  L \otimes e_{1,4} \Upsilon^{*}_{\nu}\\
&+e_{1,1}\otimes Q \otimes e_{4,1} \Upsilon_{u}\otimes e_{1,2}\otimes Q \otimes e_{1,4} \Upsilon^{*}_{u}\\
&+e_{2,1}\otimes  L \otimes e_{4,1} \Upsilon_{\nu}\otimes e_{1,1}\otimes  L \otimes e_{1,4} \Upsilon^{*}_{\nu}\\
&+e_{2,1}\otimes  L \otimes e_{4,1} \Upsilon_{\nu}\otimes e_{1,1}\otimes Q \otimes e_{1,4} \Upsilon^{*}_{u}\\
&+e_{2,1}\otimes Q \otimes e_{4,1} \Upsilon_{u}\otimes e_{1,1}\otimes  L \otimes e_{1,4} \Upsilon^{*}_{\nu}\\
&+e_{2,1}\otimes Q \otimes e_{4,1} \Upsilon_{u}\otimes e_{1,1}\otimes Q \otimes e_{1,4} \Upsilon^{*}_{u}\\
&+e_{1,2}\otimes  L \otimes e_{4,1} \Upsilon_{e}\otimes e_{2,2}\otimes  L \otimes e_{1,4} \Upsilon^{*}_{e}\\
&+e_{1,2}\otimes  L \otimes e_{4,1} \Upsilon_{e}\otimes e_{2,2}\otimes Q \otimes e_{1,4} \Upsilon^{*}_{d}\\
&+e_{1,2}\otimes Q \otimes e_{4,1} \Upsilon_{d}\otimes e_{2,2}\otimes  L \otimes e_{1,4} \Upsilon^{*}_{e}\\
&+e_{1,2}\otimes Q \otimes e_{4,1} \Upsilon_{d}\otimes e_{2,2}\otimes Q \otimes e_{1,4} \Upsilon^{*}_{d}\\
&+e_{2,2}\otimes  L \otimes e_{4,1} \Upsilon_{e}\otimes e_{2,1}\otimes  L \otimes e_{1,4} \Upsilon^{*}_{e}\\
&+e_{2,2}\otimes  L \otimes e_{4,1} \Upsilon_{e}\otimes e_{2,1}\otimes Q \otimes e_{1,4} \Upsilon^{*}_{d}\\
&+e_{2,2}\otimes Q \otimes e_{4,1} \Upsilon_{d}\otimes e_{2,1}\otimes  L \otimes e_{1,4} \Upsilon^{*}_{e}\\
&+e_{2,2}\otimes Q \otimes e_{4,1} \Upsilon_{d}\otimes e_{2,1}\otimes Q \otimes e_{1,4} \Upsilon^{*}_{d}\;,\\
\end{aligned}$}}
\\
\vtop{\null\hbox{$\begin{aligned}
h_{3} = 
&+ie_{1,1}\otimes  L \otimes e_{4,1} \Upsilon_{\nu}\otimes e_{1,2}\otimes  L \otimes e_{1,4} \Upsilon^{*}_{\nu}\\
&+ie_{1,1}\otimes  L \otimes e_{4,1} \Upsilon_{\nu}\otimes e_{1,2}\otimes Q \otimes e_{1,4} \Upsilon^{*}_{u}\\
&+ie_{1,1}\otimes Q \otimes e_{4,1} \Upsilon_{u}\otimes e_{1,2}\otimes  L \otimes e_{1,4} \Upsilon^{*}_{\nu}\\
&+ie_{1,1}\otimes Q \otimes e_{4,1} \Upsilon_{u}\otimes e_{1,2}\otimes Q \otimes e_{1,4} \Upsilon^{*}_{u}\\
&- ie_{2,1}\otimes  L \otimes e_{4,1} \Upsilon_{\nu}\otimes e_{1,1}\otimes  L \otimes e_{1,4} \Upsilon^{*}_{\nu}\\
&- ie_{2,1}\otimes  L \otimes e_{4,1} \Upsilon_{\nu}\otimes e_{1,1}\otimes Q \otimes e_{1,4} \Upsilon^{*}_{u}\\
&- ie_{2,1}\otimes Q \otimes e_{4,1} \Upsilon_{u}\otimes e_{1,1}\otimes  L \otimes e_{1,4} \Upsilon^{*}_{\nu}\\
&- ie_{2,1}\otimes Q \otimes e_{4,1} \Upsilon_{u}\otimes e_{1,1}\otimes Q \otimes e_{1,4} \Upsilon^{*}_{u}\\
&+ie_{1,2}\otimes  L \otimes e_{4,1} \Upsilon_{e}\otimes e_{2,2}\otimes  L \otimes e_{1,4} \Upsilon^{*}_{e}\\
&+ie_{1,2}\otimes  L \otimes e_{4,1} \Upsilon_{e}\otimes e_{2,2}\otimes Q \otimes e_{1,4} \Upsilon^{*}_{d}\\
&+ie_{1,2}\otimes Q \otimes e_{4,1} \Upsilon_{d}\otimes e_{2,2}\otimes  L \otimes e_{1,4} \Upsilon^{*}_{e}\\
&+ie_{1,2}\otimes Q \otimes e_{4,1} \Upsilon_{d}\otimes e_{2,2}\otimes Q \otimes e_{1,4} \Upsilon^{*}_{d}\\
&- ie_{2,2}\otimes  L \otimes e_{4,1} \Upsilon_{e}\otimes e_{2,1}\otimes  L \otimes e_{1,4} \Upsilon^{*}_{e}\\
&- ie_{2,2}\otimes  L \otimes e_{4,1} \Upsilon_{e}\otimes e_{2,1}\otimes Q \otimes e_{1,4} \Upsilon^{*}_{d}\\
&- ie_{2,2}\otimes Q \otimes e_{4,1} \Upsilon_{d}\otimes e_{2,1}\otimes  L \otimes e_{1,4} \Upsilon^{*}_{e}\\
&- ie_{2,2}\otimes Q \otimes e_{4,1} \Upsilon_{d}\otimes e_{2,1}\otimes Q \otimes e_{1,4} \Upsilon^{*}_{d}\;,\\
\end{aligned}$}}
&
\vtop{\null\hbox{$\begin{aligned}
h_{4} = 
&+e_{2,1}\otimes  L \otimes e_{4,1} \Upsilon_{\nu}\otimes e_{1,2}\otimes  L \otimes e_{1,4} \Upsilon^{*}_{\nu}\\
&+e_{2,1}\otimes  L \otimes e_{4,1} \Upsilon_{\nu}\otimes e_{1,2}\otimes Q \otimes e_{1,4} \Upsilon^{*}_{u}\\
&+e_{2,1}\otimes Q \otimes e_{4,1} \Upsilon_{u}\otimes e_{1,2}\otimes  L \otimes e_{1,4} \Upsilon^{*}_{\nu}\\
&+e_{2,1}\otimes Q \otimes e_{4,1} \Upsilon_{u}\otimes e_{1,2}\otimes Q \otimes e_{1,4} \Upsilon^{*}_{u}\\
&+e_{2,2}\otimes  L \otimes e_{4,1} \Upsilon_{e}\otimes e_{2,2}\otimes  L \otimes e_{1,4} \Upsilon^{*}_{e}\\
&+e_{2,2}\otimes  L \otimes e_{4,1} \Upsilon_{e}\otimes e_{2,2}\otimes Q \otimes e_{1,4} \Upsilon^{*}_{d}\\
&+e_{2,2}\otimes Q \otimes e_{4,1} \Upsilon_{d}\otimes e_{2,2}\otimes  L \otimes e_{1,4} \Upsilon^{*}_{e}\\
&+e_{2,2}\otimes Q \otimes e_{4,1} \Upsilon_{d}\otimes e_{2,2}\otimes Q \otimes e_{1,4} \Upsilon^{*}_{d}\;,\\
\end{aligned}$}}
\\
\end{array}
\end{equation}
\begin{equation}
\begin{array}{ll}
\vtop{\null\hbox{$\begin{aligned}
h_{5} = 
&+e_{1,1}\otimes  L \otimes e_{1,4} \Upsilon^{*}_{\nu}\otimes e_{1,1}\otimes  L \otimes e_{4,1} \Upsilon_{\nu}\\
&+e_{1,1}\otimes  L \otimes e_{1,4} \Upsilon^{*}_{\nu}\otimes e_{1,1}\otimes Q \otimes e_{4,1} \Upsilon_{u}\\
&+e_{1,1}\otimes Q \otimes e_{1,4} \Upsilon^{*}_{u}\otimes e_{1,1}\otimes  L \otimes e_{4,1} \Upsilon_{\nu}\\
&+e_{1,1}\otimes Q \otimes e_{1,4} \Upsilon^{*}_{u}\otimes e_{1,1}\otimes Q \otimes e_{4,1} \Upsilon_{u}\\
&+e_{1,2}\otimes  L \otimes e_{1,4} \Upsilon^{*}_{\nu}\otimes e_{2,1}\otimes  L \otimes e_{4,1} \Upsilon_{\nu}\\
&+e_{1,2}\otimes  L \otimes e_{1,4} \Upsilon^{*}_{\nu}\otimes e_{2,1}\otimes Q \otimes e_{4,1} \Upsilon_{u}\\
&+e_{1,2}\otimes Q \otimes e_{1,4} \Upsilon^{*}_{u}\otimes e_{2,1}\otimes  L \otimes e_{4,1} \Upsilon_{\nu}\\
&+e_{1,2}\otimes Q \otimes e_{1,4} \Upsilon^{*}_{u}\otimes e_{2,1}\otimes Q \otimes e_{4,1} \Upsilon_{u}\;,\\
\end{aligned}$}}
&
\vtop{\null\hbox{$\begin{aligned}
h_{6} = 
&+e_{1,1}\otimes  L \otimes e_{1,4} \Upsilon^{*}_{\nu}\otimes e_{1,2}\otimes  L \otimes e_{4,1} \Upsilon_{e}\\
&+e_{1,1}\otimes  L \otimes e_{1,4} \Upsilon^{*}_{\nu}\otimes e_{1,2}\otimes Q \otimes e_{4,1} \Upsilon_{d}\\
&+e_{1,1}\otimes Q \otimes e_{1,4} \Upsilon^{*}_{u}\otimes e_{1,2}\otimes  L \otimes e_{4,1} \Upsilon_{e}\\
&+e_{1,1}\otimes Q \otimes e_{1,4} \Upsilon^{*}_{u}\otimes e_{1,2}\otimes Q \otimes e_{4,1} \Upsilon_{d}\\
&+e_{1,2}\otimes  L \otimes e_{1,4} \Upsilon^{*}_{\nu}\otimes e_{2,2}\otimes  L \otimes e_{4,1} \Upsilon_{e}\\
&+e_{1,2}\otimes  L \otimes e_{1,4} \Upsilon^{*}_{\nu}\otimes e_{2,2}\otimes Q \otimes e_{4,1} \Upsilon_{d}\\
&+e_{1,2}\otimes Q \otimes e_{1,4} \Upsilon^{*}_{u}\otimes e_{2,2}\otimes  L \otimes e_{4,1} \Upsilon_{e}\\
&+e_{1,2}\otimes Q \otimes e_{1,4} \Upsilon^{*}_{u}\otimes e_{2,2}\otimes Q \otimes e_{4,1} \Upsilon_{d}\\
&+e_{2,1}\otimes  L \otimes e_{1,4} \Upsilon^{*}_{e}\otimes e_{1,1}\otimes  L \otimes e_{4,1} \Upsilon_{\nu}\\
&+e_{2,1}\otimes  L \otimes e_{1,4} \Upsilon^{*}_{e}\otimes e_{1,1}\otimes Q \otimes e_{4,1} \Upsilon_{u}\\
&+e_{2,1}\otimes Q \otimes e_{1,4} \Upsilon^{*}_{d}\otimes e_{1,1}\otimes  L \otimes e_{4,1} \Upsilon_{\nu}\\
&+e_{2,1}\otimes Q \otimes e_{1,4} \Upsilon^{*}_{d}\otimes e_{1,1}\otimes Q \otimes e_{4,1} \Upsilon_{u}\\
&+e_{2,2}\otimes  L \otimes e_{1,4} \Upsilon^{*}_{e}\otimes e_{2,1}\otimes  L \otimes e_{4,1} \Upsilon_{\nu}\\
&+e_{2,2}\otimes  L \otimes e_{1,4} \Upsilon^{*}_{e}\otimes e_{2,1}\otimes Q \otimes e_{4,1} \Upsilon_{u}\\
&+e_{2,2}\otimes Q \otimes e_{1,4} \Upsilon^{*}_{d}\otimes e_{2,1}\otimes  L \otimes e_{4,1} \Upsilon_{\nu}\\
&+e_{2,2}\otimes Q \otimes e_{1,4} \Upsilon^{*}_{d}\otimes e_{2,1}\otimes Q \otimes e_{4,1} \Upsilon_{u}\;,\\
\end{aligned}$}}
\\
\vtop{\null\hbox{$\begin{aligned}
h_{7} = 
&+ie_{1,1}\otimes  L \otimes e_{1,4} \Upsilon^{*}_{\nu}\otimes e_{1,2}\otimes  L \otimes e_{4,1} \Upsilon_{e}\\
&+ie_{1,1}\otimes  L \otimes e_{1,4} \Upsilon^{*}_{\nu}\otimes e_{1,2}\otimes Q \otimes e_{4,1} \Upsilon_{d}\\
&+ie_{1,1}\otimes Q \otimes e_{1,4} \Upsilon^{*}_{u}\otimes e_{1,2}\otimes  L \otimes e_{4,1} \Upsilon_{e}\\
&+ie_{1,1}\otimes Q \otimes e_{1,4} \Upsilon^{*}_{u}\otimes e_{1,2}\otimes Q \otimes e_{4,1} \Upsilon_{d}\\
&+ie_{1,2}\otimes  L \otimes e_{1,4} \Upsilon^{*}_{\nu}\otimes e_{2,2}\otimes  L \otimes e_{4,1} \Upsilon_{e}\\
&+ie_{1,2}\otimes  L \otimes e_{1,4} \Upsilon^{*}_{\nu}\otimes e_{2,2}\otimes Q \otimes e_{4,1} \Upsilon_{d}\\
&+ie_{1,2}\otimes Q \otimes e_{1,4} \Upsilon^{*}_{u}\otimes e_{2,2}\otimes  L \otimes e_{4,1} \Upsilon_{e}\\
&+ie_{1,2}\otimes Q \otimes e_{1,4} \Upsilon^{*}_{u}\otimes e_{2,2}\otimes Q \otimes e_{4,1} \Upsilon_{d}\\
&- ie_{2,1}\otimes  L \otimes e_{1,4} \Upsilon^{*}_{e}\otimes e_{1,1}\otimes  L \otimes e_{4,1} \Upsilon_{\nu}\\
&- ie_{2,1}\otimes  L \otimes e_{1,4} \Upsilon^{*}_{e}\otimes e_{1,1}\otimes Q \otimes e_{4,1} \Upsilon_{u}\\
&- ie_{2,1}\otimes Q \otimes e_{1,4} \Upsilon^{*}_{d}\otimes e_{1,1}\otimes  L \otimes e_{4,1} \Upsilon_{\nu}\\
&- ie_{2,1}\otimes Q \otimes e_{1,4} \Upsilon^{*}_{d}\otimes e_{1,1}\otimes Q \otimes e_{4,1} \Upsilon_{u}\\
&- ie_{2,2}\otimes  L \otimes e_{1,4} \Upsilon^{*}_{e}\otimes e_{2,1}\otimes  L \otimes e_{4,1} \Upsilon_{\nu}\\
&- ie_{2,2}\otimes  L \otimes e_{1,4} \Upsilon^{*}_{e}\otimes e_{2,1}\otimes Q \otimes e_{4,1} \Upsilon_{u}\\
&- ie_{2,2}\otimes Q \otimes e_{1,4} \Upsilon^{*}_{d}\otimes e_{2,1}\otimes  L \otimes e_{4,1} \Upsilon_{\nu}\\
&- ie_{2,2}\otimes Q \otimes e_{1,4} \Upsilon^{*}_{d}\otimes e_{2,1}\otimes Q \otimes e_{4,1} \Upsilon_{u}\;,\\
\end{aligned}$}}
&
\vtop{\null\hbox{$\begin{aligned}
h_{8} = 
&+e_{2,1}\otimes  L \otimes e_{1,4} \Upsilon^{*}_{e}\otimes e_{1,2}\otimes  L \otimes e_{4,1} \Upsilon_{e}\\
&+e_{2,1}\otimes  L \otimes e_{1,4} \Upsilon^{*}_{e}\otimes e_{1,2}\otimes Q \otimes e_{4,1} \Upsilon_{d}\\
&+e_{2,1}\otimes Q \otimes e_{1,4} \Upsilon^{*}_{d}\otimes e_{1,2}\otimes  L \otimes e_{4,1} \Upsilon_{e}\\
&+e_{2,1}\otimes Q \otimes e_{1,4} \Upsilon^{*}_{d}\otimes e_{1,2}\otimes Q \otimes e_{4,1} \Upsilon_{d}\\
&+e_{2,2}\otimes  L \otimes e_{1,4} \Upsilon^{*}_{e}\otimes e_{2,2}\otimes  L \otimes e_{4,1} \Upsilon_{e}\\
&+e_{2,2}\otimes  L \otimes e_{1,4} \Upsilon^{*}_{e}\otimes e_{2,2}\otimes Q \otimes e_{4,1} \Upsilon_{d}\\
&+e_{2,2}\otimes Q \otimes e_{1,4} \Upsilon^{*}_{d}\otimes e_{2,2}\otimes  L \otimes e_{4,1} \Upsilon_{e}\\
&+e_{2,2}\otimes Q \otimes e_{1,4} \Upsilon^{*}_{d}\otimes e_{2,2}\otimes Q \otimes e_{4,1} \Upsilon_{d}\;.\\
\end{aligned}$}}
\\
\end{array}
\end{equation}
The final ingredient required to define a space $\Lambda^2_D$ is an idempotent bimodule map $\Psi$ from $T^2_D$ to itself whose range is $JT^2_D$. Recall that this choice of $\Psi$ need not be unique. We present two candidates, $\Psi$ and $\Phi$.\\
We define $\Psi$ by setting $\Psi(\frak{h}_j) = \frak{h}_j$ and $\Psi(h_i) = 0$. Note that $\Psi$ maps Hermitian elements to Hermitian elements, and is thus compatible with the involution of the ambient space $B(\mathcal{H})$.\\
Our second candidate $\Phi$ is more involved. The justification for this map will be given in Section \ref{16sep25sm4}. $\Phi$ is of course identity on $\frak{h}$. We define $\Phi$ going from the ordered sub--basis $h_i$ to the ordered sub--basis $\frak{h}_j$ by the following matrix.
\begin{equation}
\label{phi h}
\Phi\rvert_{h} = \left[\begin{matrix}1 & 0 & 0 & 0 & 0 & 0 & 0 & 0\\0 & 0 & - i & 0 & 0 & 0 & 0 & 0\\0 & i & 0 & 0 & 0 & 0 & 0 & 0\\0 & 0 & 0 & -1 & 0 & 0 & 0 & 0\end{matrix}\right]
\end{equation}
It is easy to see that $\Phi$ is an idempotent map. For example, $\Phi^2(h_1) = \Phi(\frak{h}_1) = \frak{h}_1$. Moreover, due to the appearance of complex entries, this map is not a Hermitian map, and thus not compatible with the involution. Finally, both $\Psi$ and $\Phi$ are bimodule maps, i.e., they are left and right $A_F$-linear. This is not trivial to see for $\Phi$, but it is true. One can verify for arbitrary $a\in A_F$ and $x\in T^2_D$ that
$$\Phi(ax) = a \Phi(x)\;,\quad \Phi(xa) = \Phi(x)a\;.$$\\

Thus we have given two non--equivalent second order differential calculi based on the Mesland--Rennie formalism, which we denote by $\Lambda^2_\Psi$ and $\Lambda^2_\Phi$. The exterior derivatives $\dee_\Psi$ and $\dee_\Phi$ are obtained via \eqref{16sep25sm5}.

\section{Algebraic functionals} \label{16sep25sm4}

In Section \ref{16sep25sm6}, we discussed spectral metric functional and spectral torsion functional. In this section, we discuss algebraic analogues for the same, viz. describe algebra valued maps, which when post--composed by the trace on the space $B(\mathcal{H})$ recovers the same information as obtained from the purely spectral data.

\subsection{Algebraic metric functional}

For our finite geometry, the role of the Wodzicki residue (as well as noncommutative integral and Dixmier trace $\Tr^+$) is played by the trace on $B(\mathcal{H})$. In the language of Hilbert inner products, our $\IC$--bilinear spectral metric functional is thus given by the formula
\begin{equation}
    g(u,v) = {}_\IC \langle u, v^* \rangle = \Tr(uv).
\end{equation}
In \cite[Theorem 2.9]{FGR99}, it was proven that given the Hilbert inner product ${}_\IC \langle \cdot, \cdot \rangle$, there exists a well--defined sesquilinear map, called a generalised Hermitian structure,
\begin{equation}
    {}_A \langle \cdot, \cdot \rangle : \Omega^1_D \times \Omega^1_D \to \overline{A}\;,
\end{equation}
taking values in the von Neumann algebra $\overline{A}$ generated by the algebra $A$, which satisfies the relation
\begin{equation}
    \text{Tr}^+\left({}_A \langle \cdot, \cdot \rangle \right)= {}_\IC \langle \cdot, \cdot \rangle\;.
\end{equation}
Our algebra is finite dimensional, hence $\overline{A} = A$. Thus, we have an $A$--bilinear map
\begin{equation}
    g_A(u,v) = {}_A \langle u, v^\ast \rangle,
\end{equation}
which gives us an algebra valued metric functional satisfying
\begin{equation}
    g(u,v) = \Tr\left( g_A(u,v) \right)\;.
\end{equation}
As an example, on our space of 1--forms $\Omega^1_D$, the unique Hermitian structure which reproduces $g$ on the four generating elements $\omega_{\alpha - z, \beta, \gamma, \delta - w}$ is given by
\begin{equation}
\begin{split}
\;_A\langle \omega_{\alpha - z}, \omega_{\alpha - z}\rangle =
&+\left[\begin{matrix}\frac{\Tr{\left(\Upsilon^{*}_{\nu} \Upsilon_{\nu} \right)}}{8} + \frac{3 \Tr{\left(\Upsilon^{*}_{u} \Upsilon_{u} \right)}}{8} & 0\\0 & 0\end{matrix}\right]\otimes\left[\begin{matrix}1 & 0 & 0 & 0\\0 & 1 & 0 & 0\\0 & 0 & 1 & 0\\0 & 0 & 0 & 1\end{matrix}\right]\otimes\left[\begin{matrix}1 & 0 & 0 & 0\\0 & 0 & 0 & 0\\0 & 0 & 0 & 0\\0 & 0 & 0 & 0\end{matrix}\right]\otimes \mathds{1}_3\\
&+\left[\begin{matrix}\frac{\Tr{\left(\Upsilon^{*}_{\nu} \Upsilon_{\nu} \right)}}{8} + \frac{3 \Tr{\left(\Upsilon^{*}_{u} \Upsilon_{u} \right)}}{8} & 0\\0 & \frac{\Tr{\left(\Upsilon^{*}_{\nu} \Upsilon_{\nu} \right)}}{8} + \frac{3 \Tr{\left(\Upsilon^{*}_{u} \Upsilon_{u} \right)}}{8}\end{matrix}\right]\otimes\left[\begin{matrix}1 & 0 & 0 & 0\\0 & 0 & 0 & 0\\0 & 0 & 0 & 0\\0 & 0 & 0 & 0\end{matrix}\right]\otimes\left[\begin{matrix}0 & 0 & 0 & 0\\0 & 1 & 0 & 0\\0 & 0 & 1 & 0\\0 & 0 & 0 & 0\end{matrix}\right]\otimes \mathds{1}_3\\
&+\left[\begin{matrix}\frac{\Tr{\left(\Upsilon^{*}_{\nu} \Upsilon_{\nu} \right)}}{4} + \frac{3 \Tr{\left(\Upsilon^{*}_{u} \Upsilon_{u} \right)}}{4} & 0\\0 & 0\end{matrix}\right]\otimes\left[\begin{matrix}1 & 0 & 0 & 0\\0 & 1 & 0 & 0\\0 & 0 & 1 & 0\\0 & 0 & 0 & 1\end{matrix}\right]\otimes\left[\begin{matrix}0 & 0 & 0 & 0\\0 & 0 & 0 & 0\\0 & 0 & 0 & 0\\0 & 0 & 0 & 1\end{matrix}\right]\otimes \mathds{1}_3\;,\\
\;_A\langle \omega_{\gamma}, \omega_{\gamma}\rangle =
&+\left[\begin{matrix}0 & 0\\0 & \frac{3 \Tr{\left(\Upsilon^{*}_{d} \Upsilon_{d} \right)}}{4} + \frac{\Tr{\left(\Upsilon^{*}_{e} \Upsilon_{e} \right)}}{4}\end{matrix}\right]\otimes\left[\begin{matrix}1 & 0 & 0 & 0\\0 & 1 & 0 & 0\\0 & 0 & 1 & 0\\0 & 0 & 0 & 1\end{matrix}\right]\otimes\left[\begin{matrix}1 & 0 & 0 & 0\\0 & 0 & 0 & 0\\0 & 0 & 0 & 0\\0 & 0 & 0 & 0\end{matrix}\right]\otimes \mathds{1}_3\\
&+\left[\begin{matrix}0 & 0\\0 & \frac{\Tr{\left(\Upsilon^{*}_{\nu} \Upsilon_{\nu} \right)}}{4} + \frac{3 \Tr{\left(\Upsilon^{*}_{u} \Upsilon_{u} \right)}}{4}\end{matrix}\right]\otimes\left[\begin{matrix}1 & 0 & 0 & 0\\0 & 1 & 0 & 0\\0 & 0 & 1 & 0\\0 & 0 & 0 & 1\end{matrix}\right]\otimes\left[\begin{matrix}0 & 0 & 0 & 0\\0 & 0 & 0 & 0\\0 & 0 & 0 & 0\\0 & 0 & 0 & 1\end{matrix}\right]\otimes \mathds{1}_3\;,\\
\end{split}
\end{equation}
\begin{equation}
\begin{split}
\;_A\langle \omega_{\beta}, \omega_{\beta}\rangle =
&+\left[\begin{matrix}\frac{\Tr{\left(\Upsilon^{*}_{\nu} \Upsilon_{\nu} \right)}}{8} + \frac{3 \Tr{\left(\Upsilon^{*}_{u} \Upsilon_{u} \right)}}{8} & 0\\0 & 0\end{matrix}\right]\otimes\left[\begin{matrix}1 & 0 & 0 & 0\\0 & 1 & 0 & 0\\0 & 0 & 1 & 0\\0 & 0 & 0 & 1\end{matrix}\right]\otimes\left[\begin{matrix}1 & 0 & 0 & 0\\0 & 0 & 0 & 0\\0 & 0 & 0 & 0\\0 & 0 & 0 & 0\end{matrix}\right]\otimes \mathds{1}_3\\
&+\left[\begin{matrix}\frac{\Tr{\left(\Upsilon^{*}_{\nu} \Upsilon_{\nu} \right)}}{8} + \frac{3 \Tr{\left(\Upsilon^{*}_{u} \Upsilon_{u} \right)}}{8} & 0\\0 & \frac{\Tr{\left(\Upsilon^{*}_{\nu} \Upsilon_{\nu} \right)}}{8} + \frac{3 \Tr{\left(\Upsilon^{*}_{u} \Upsilon_{u} \right)}}{8}\end{matrix}\right]\otimes\left[\begin{matrix}1 & 0 & 0 & 0\\0 & 0 & 0 & 0\\0 & 0 & 0 & 0\\0 & 0 & 0 & 0\end{matrix}\right]\otimes\left[\begin{matrix}0 & 0 & 0 & 0\\0 & 1 & 0 & 0\\0 & 0 & 1 & 0\\0 & 0 & 0 & 0\end{matrix}\right]\otimes \mathds{1}_3\\
&+\left[\begin{matrix}\frac{3 \Tr{\left(\Upsilon^{*}_{d} \Upsilon_{d} \right)}}{4} + \frac{\Tr{\left(\Upsilon^{*}_{e} \Upsilon_{e} \right)}}{4} & 0\\0 & 0\end{matrix}\right]\otimes\left[\begin{matrix}1 & 0 & 0 & 0\\0 & 1 & 0 & 0\\0 & 0 & 1 & 0\\0 & 0 & 0 & 1\end{matrix}\right]\otimes\left[\begin{matrix}0 & 0 & 0 & 0\\0 & 0 & 0 & 0\\0 & 0 & 0 & 0\\0 & 0 & 0 & 1\end{matrix}\right]\otimes \mathds{1}_3\;,\\
\;_A\langle \omega_{\delta - w}, \omega_{\delta - w}\rangle =
&+\left[\begin{matrix}0 & 0\\0 & \frac{3 \Tr{\left(\Upsilon^{*}_{d} \Upsilon_{d} \right)}}{4} + \frac{\Tr{\left(\Upsilon^{*}_{e} \Upsilon_{e} \right)}}{4}\end{matrix}\right]\otimes\left[\begin{matrix}1 & 0 & 0 & 0\\0 & 1 & 0 & 0\\0 & 0 & 1 & 0\\0 & 0 & 0 & 1\end{matrix}\right]\otimes\left[\begin{matrix}1 & 0 & 0 & 0\\0 & 0 & 0 & 0\\0 & 0 & 0 & 0\\0 & 0 & 0 & 0\end{matrix}\right]\otimes \mathds{1}_3\\
&+\left[\begin{matrix}0 & 0\\0 & \frac{3 \Tr{\left(\Upsilon^{*}_{d} \Upsilon_{d} \right)}}{4} + \frac{\Tr{\left(\Upsilon^{*}_{e} \Upsilon_{e} \right)}}{4}\end{matrix}\right]\otimes\left[\begin{matrix}1 & 0 & 0 & 0\\0 & 1 & 0 & 0\\0 & 0 & 1 & 0\\0 & 0 & 0 & 1\end{matrix}\right]\otimes\left[\begin{matrix}0 & 0 & 0 & 0\\0 & 0 & 0 & 0\\0 & 0 & 0 & 0\\0 & 0 & 0 & 1\end{matrix}\right]\otimes \mathds{1}_3\;,\\
\;_A\langle \omega_{\gamma}, \omega_{\alpha - z}\rangle =&+\left[\begin{matrix}0 & 0\\\frac{\Tr{\left(\Upsilon^{*}_{\nu} \Upsilon_{\nu} \right)}}{4} + \frac{3 \Tr{\left(\Upsilon^{*}_{u} \Upsilon_{u} \right)}}{4} & 0\end{matrix}\right]\otimes\left[\begin{matrix}1 & 0 & 0 & 0\\0 & 1 & 0 & 0\\0 & 0 & 1 & 0\\0 & 0 & 0 & 1\end{matrix}\right]\otimes\left[\begin{matrix}0 & 0 & 0 & 0\\0 & 0 & 0 & 0\\0 & 0 & 0 & 0\\0 & 0 & 0 & 1\end{matrix}\right]\otimes \mathds{1}_3\;,\\
\;_A\langle \omega_{\beta}, \omega_{\delta - w}\rangle =&+\left[\begin{matrix}0 &  \frac{3 \Tr{\left(\Upsilon^{*}_{d} \Upsilon_{d} \right)}}{4}  +\frac{\Tr{\left(\Upsilon^{*}_{e} \Upsilon_{e} \right)}}{4}\\0 & 0\end{matrix}\right]\otimes\left[\begin{matrix}1 & 0 & 0 & 0\\0 & 1 & 0 & 0\\0 & 0 & 1 & 0\\0 & 0 & 0 & 1\end{matrix}\right]\otimes\left[\begin{matrix}0 & 0 & 0 & 0\\0 & 0 & 0 & 0\\0 & 0 & 0 & 0\\0 & 0 & 0 & 1\end{matrix}\right]\otimes \mathds{1}_3\;,\\
\end{split}
\end{equation}
where the other expression are determined by skew--symmetry or are equal to zero.

\subsection{Algebraic torsion functional}\label{VIB}
Given a second order differential calculus $(\Omega^1(A)$, $T^2(A) = \Omega^1(A)\otimes_A\Omega^1(A)$, $ \dee:\Omega^1(A)\rightarrow T^2(A))$ 
on a (non--commutative) algebra $A$, and a (left) connection
\[ \nabla : \Omega^1(A) \to \Omega^1(A) \otimes_A \Omega^1(A), \]
algebraic torsion of the connection $T^\nabla$ is the left $A$--linear map
\begin{equation}
    \begin{array}{cc}
         &  T^\nabla : \Omega^1(A) \to \Omega^1(A)\otimes_A \Omega^1(A);\\
         &  T^\nabla(\omega) = \nabla (\omega) - \dee(\omega).
    \end{array}
\label{torsion left linear map}
\end{equation}
The torsion map depends on the construction of the second order differential calculus and is well studied in literature and we refer to \cite{beggs2020quantum} 
for details.\\
In case the differential calculus comes form a finitely summable and smooth spectral triple which admits a generalized Wodzicki residue 
(or equivalently Dixmier trace $\text{Tr}^+$) the following  algebraic torsion functional is defined as in \cite[Definition 2.11]{dkabrowski2024algebraic} by
\begin{equation}
    \mathcal{T}(u,v,w) = \text{Tr}^+\left(uv\;\mu\circ T^\nabla(w)\right)\;.
\label{algebraic torsion functional general}
\end{equation}
Here, $u,v,w\in \Omega^1_D(A)$ and $\mu : \Omega^1_D(A) \otimes_A \Omega^1_D(A) \to \Omega^2_D(A)$, is the multiplication map 
of the differential graded algebra.\\
 We will study 
\eqref{algebraic torsion functional general}
for the finite geometry \eqref{complexified SM} (so $\text{Tr}^+=\text{Tr}$) both in the Connes and the Mesland--Rennie formalisms, 
and compare with the spectral torsion functional. Note however, that any algebraic notion of torsion necessarily involves a connection, 
while the spectral notion is connection agnostic. Thus our task is to search for a (possibly unique) connection on the second order differetial calculi 
which is able to recover the purely spectral data. We will show that such a connection does not exist in the Connes formalism, 
but it does in the Mesland-Rennie formalism.
\subsubsection*{Torsion for the Connes formalism}
The ingredients from the aformentioned introduction to torsion can be translated to Connes' formalism as follows. Simply set $\Omega^1(A) = \Omega^1_D$ and $\Omega^2(A) = \Omega^2_D$. Regarding the differential, since the 2--forms $\Omega^2_D$ are defined as $\pi_D(\Omega^2_u)$ (orthogonally) quotiented by $J^2_D$, the differential on 1--forms $\dee:\Omega^1_D\rightarrow \Omega^2_D$
is a well defined map
$$\dee (\sum_j a_j\left[D,b_j\right]) = \sum_j \left[D,a_j\right]\left[D,b_j\right] $$
in the sense that it does not depend on the decomposition of the 1--form into $\sum_j a_j \left[D,b_j\right]$. This differential map $\dee$ takes place of the $\mu\circ\dee$ from $\mu\circ T^\nabla$ in \eqref{algebraic torsion functional general}.
Next, instead of considering tensor valued connections $\nabla$ and then applying the multiplication map, for our negative result it is enough to try to search for the composed map 
$$\mu\circ\nabla : \Omega^1_D \rightarrow \Omega^2_D\;,$$
implicit in \eqref{torsion left linear map}, which solves 
\begin{equation}
    \Tr (uvwD) = \Tr(uv\,\mu\circ T^\nabla(w))
\label{algebraic torsion functional}
\end{equation}
for all $u,v,w\in \Omega^1_D$. We refer to the RHS of \eqref{algebraic torsion functional} as the \textit{algebraic torsion functional} in the Connes formalism. Since $T^\nabla$ is left $A_F$-linear, and $uv\in\pi_D(\Omega^2_u)$ which is an $A_F$-bimodule, it is enough to focus on the equality
\begin{equation}
\label{uvdaD = uvTda}
    \Tr(uv\dee(a)D) = \Tr(uv\,\mu\circ T^\nabla(\dee(a)))
\end{equation}
for general $u,v,\in\Omega^1_D$ and $a\in A_F$. Now it is clear that it is enough to find the action of $\mu\circ\nabla$ on the generators of 1--forms $\omega_{\alpha-z},\omega_\beta,\omega_\gamma,\omega_{\delta-w}$ which solves \eqref{uvdaD = uvTda}. We adopt the most general Ansatz by expanding each $\mu\circ\nabla(\omega_{...})$ as a general sum of the $8$ twoforms \eqref{C twoforms basis} multiplied with arbitrary complex coefficients as follows
\begin{equation}
\begin{split}
\nabla(\omega_{\alpha-z}) &= \sum_i C^i_{\alpha - z} \;\lambda_i\;,\quad \forall C^i_{\alpha-z}\in \mathbb{C}, \\
\nabla(\omega_{\beta}) &= \sum_i C^i_{\beta} \;\lambda_i\;,\quad \forall C^i_{\beta}\in \mathbb{C}, \\
\nabla(\omega_{\gamma}) &= \sum_i\; C^i_{\gamma} \lambda_i\;,\quad \forall C^i_{\gamma}\in \mathbb{C}, \\
\nabla(\omega_{\delta-w}) &= \sum_i C^i_{\delta - w} \;\lambda_i\;,\quad \forall C^i_{\delta-w}\in \mathbb{C}.\\
\end{split}
\label{Connes torsion ansatz}
\end{equation}
This turns \eqref{uvdaD = uvTda} into a system of linear equations in the $4\times 8$ variables\footnote{The product $uv$ lies in $\pi_D(\Omega^2_u)$ and as such can be expanded as $uv = \sum_i D_i \Omega_i$ for parameters $D_i\in \IC$ and $\Omega_i$ the basis elements from \eqref{pi_D Omega^2_u}. Additionally, as shown in \eqref{da_psi}, any exact oneform can be expanded as $\dee(a) = (\alpha-z)\omega_{\alpha-z} + ... + (\delta-w)\omega_{\delta-w}$. Then, \eqref{uvdaD = uvTda} produces an equation in the $4\times 8$ Ansatz's variables for every distinct product of $D_i$ and $(\alpha-z),\beta,\gamma,(\delta-w)$.}. However, it turns out that this system does not have a solution in the case of Standard Model's internal NC geometry. In other words, the Connes formalism in the way it was presented in this paper can not equate the algebraic and spectral torsion functionals of the internal part of the Standard Model's spectral triple.
\subsubsection*{Torsion for the Mesland-Rennie formalism}

In \cite[Section 4]{dkabrowski2024algebraic}, it was shown that for a 1--form $u$ in the second order differential calculus $(\Omega^1_D, \Lambda^2_D, \dee_\Psi)$ on the two sheeted manifold $M \times \Z_2$ constructed via the Mesland--Rennie formalism, one gets the relation
\begin{equation}
    wD = \mu \circ T^\nabla_\Psi(w).
\end{equation}
Thus the following comparison result with the spectral torsion functional was shown to hold trivially:
\begin{equation}
    \mathcal{T}_D(u,v,w) := \int^\W uvwD = \int^\W u v \mu (T^\nabla_\Psi(w))\;,
\end{equation}
where $T^\nabla_\Psi$ is the torsion in the sense of \eqref{torsion left linear map} applied to Mesland-Rennie calculus and it is defined as \cite{mesland2025existence,dkabrowski2024algebraic}
\begin{equation}
    T^\nabla_\Psi = (1-\Psi)\circ \nabla - \dee_\Psi\;.
\end{equation}
Informed by this result, we compute the expressions $\omega D$, for the four generating elements $\omega_{\alpha- z}, \omega_\beta, \omega_\gamma, \omega_{\delta - w}$.
\begin{equation}
\begin{array}{ll}
\vtop{\null\hbox{$\begin{aligned}
\omega_{\alpha - z}D = 
&+e_{1,1}\otimes  L \otimes \left(e_{1,1} \Upsilon^{*}_{\nu} \Upsilon_{\nu} - e_{4,3} \Upsilon_{\nu} \Upsilon_{R} - e_{4,4} \Upsilon_{\nu} \Upsilon^{*}_{\nu}\right)\\
&+e_{1,1}\otimes Q \otimes \left(e_{1,1} \Upsilon^{*}_{u} \Upsilon_{u} - e_{4,4} \Upsilon_{u} \Upsilon^{*}_{u}\right)\\
\end{aligned}$}}
&
\vtop{\null\hbox{$\begin{aligned}
\omega_{\beta}D = 
&+e_{1,2}\otimes  L \otimes \left(e_{1,1} \Upsilon^{*}_{\nu} \Upsilon_{e} - e_{4,4} \Upsilon_{e} \Upsilon^{*}_{e}\right)\\
&+e_{1,2}\otimes Q \otimes \left(e_{1,1} \Upsilon^{*}_{u} \Upsilon_{d} - e_{4,4} \Upsilon_{d} \Upsilon^{*}_{d}\right)\\
\end{aligned}$}}
\\
\vtop{\null\hbox{$\begin{aligned}
\omega_{\gamma}D = 
&+e_{2,1}\otimes  L \otimes \left(e_{1,1} \Upsilon^{*}_{e} \Upsilon_{\nu} - e_{4,3} \Upsilon_{\nu} \Upsilon_{R} - e_{4,4} \Upsilon_{\nu} \Upsilon^{*}_{\nu}\right)\\
&+e_{2,1}\otimes Q \otimes \left(e_{1,1} \Upsilon^{*}_{d} \Upsilon_{u} - e_{4,4} \Upsilon_{u} \Upsilon^{*}_{u}\right)\\
\end{aligned}$}}
&
\vtop{\null\hbox{$\begin{aligned}
\omega_{\delta - w}D = 
&+e_{2,2}\otimes  L \otimes \left(e_{1,1} \Upsilon^{*}_{e} \Upsilon_{e} - e_{4,4} \Upsilon_{e} \Upsilon^{*}_{e}\right)\\
&+e_{2,2}\otimes Q \otimes \left(e_{1,1} \Upsilon^{*}_{d} \Upsilon_{d} - e_{4,4} \Upsilon_{d} \Upsilon^{*}_{d}\right)\\
\end{aligned}$}}
\\
\end{array}
\end{equation}
The appearance of $\Upsilon_R$ in two of the four expressions suggests that we will not be able to get the result $\mu \circ T^{\nabla}_\Psi(\omega) = \omega D$ on the nose. This is because in the algebraic description of $T^\nabla_\Psi(\omega)$, $\Upsilon_R$ does not appear at all. However, it can be seen from \eqref{16sep25sm7}, that for an arbitrary element $X$ in $\pi_D(\Omega^3_u)$, the expression $Tr(XD)$ kills all components carrying the coefficient $\Upsilon_R$. Thus we search for both an idempotent $\Psi$ and a connection $\nabla$ such that $T^\nabla_\Psi(\omega) = \mu^{-1}(\omega D_F|_{\Upsilon_R=0})$.\\

The summary of this search is the following. There exists a unique idempotent bimodule map $\Psi$, given as $\Psi = \Phi$ from \eqref{phi h}, for which there exists a connection $\nabla$ which solves
\begin{equation}
\label{algebraic = spectral MR}
    \Tr(uvwD) = \Tr(uv\;\mu\circ T^\nabla_\Phi(w))
\end{equation}
for all $u,v,w\in\Omega^1_D$. A connection which allows \eqref{algebraic = spectral MR} is given as
\begin{equation}
\label{solution connection}
\begin{split}
\nabla(\omega_{\alpha - z}) &= \frac{1}{2}\left( - h_1 + h_5\right)\;, \\
\nabla(\omega_\gamma) &= \frac{1}{4}\left( -h_2 -ih_3 + h_6 + ih_7\right)\;,\\
\nabla(\omega_\beta) &= \frac{1}{4}\left( -h_2 +ih_3 + h_6 - ih_7\right)\;,\\
\nabla(\omega_{\delta - w}) &= \frac{1}{2}\left(h_4 - h_8\right)\;,
\end{split}
\end{equation}
but also any connection $\nabla'$ which is given as  \begin{equation}
    \nabla'(\omega) = \nabla(\omega) + C_1(\omega) \mathfrak{h}_1 + C_2(\omega) \mathfrak{h}_2 + C_3(\omega) \mathfrak{h}_3+ C_4(\omega) \mathfrak{h}_4\;,\\
\end{equation}
with $C_1(\omega),C_2(\omega),C_3(\omega),C_4(\omega)\in \mathbb{C}$ and $\omega = \omega_{\alpha-z},\omega_\beta,\omega_\gamma, \omega_{\delta-w}$, is also a connection which solves \eqref{algebraic = spectral MR}. If one considers the $(1-\Phi)\nabla$ as the connection, then the connection is unique and its action on 1--form generators (over $A_F$) is given as
\begin{equation}
\begin{split}
(1-\Phi)\nabla(\omega_{\alpha - z}) &= \frac{1}{2}\left(\mathfrak{h}_1 - h_1 + h_5\right)\;, \\
(1-\Phi)\nabla(\omega_\gamma) &= \frac{1}{4}\left(\mathfrak{h}_2 +i\mathfrak{h}_3 -h_2 -ih_3 + h_6 + ih_7\right)\;,\\
(1-\Phi)\nabla(\omega_\beta) &= \frac{1}{4}\left(-\mathfrak{h}_2 +i\mathfrak{h}_3 -h_2 +ih_3 + h_6 - ih_7\right)\;,\\
(1-\Phi)\nabla(\omega_{\delta - w}) &= \frac{1}{2}\left(\mathfrak{h}_4 + h_4 - h_8\right)\;.
\end{split}
\end{equation}
In that sense, the algebraic data which reproduces the spectral torsion functional is unique in the case of Standard Model's internal spectral triple. This is the second main result of this paper.
\section{Concluding remarks}
The first main result of this paper is that the internal geometry of the noncommutative Standard Model has nonvanishing torsion, codified in terms of the spectral torsion functional of quantum 1-forms. 
Since this torsion will persist after taking the product with the canonical spectral triple on a spin manifold $M$, i.e. passing to the full almost commutative geometry of the Standard Model, we can proclaim that at each point of $M$ there will be nonvanishing components of the torsion in the internal (noncommutative) directions. It will be interesting to analyze what could be the physical impact of $M$ being permeated with such a "quantum" torsion. In particular if it could be employed in relation to dark matter and energy, 
as the usual torsion tensor discussed in the literature (see e.g. \cite{Grensing:2020maa, delaCruzDombriz:2021nrg, Pereira:2022cmu}). Additionally, it would be interesting to see the impact of our internal torsion on the torsional alterative to cosmological inflation \cite{Poplawski:2010kb} or if we managed to provide a source of torsion to the torsion-curvature duality from \cite{Dzhunushaliev:2012vb}.

The second main result is that this functional agrees with an appropriate functional defined in terms of a connection 
on the fist order differential calculus, provided we employ its recent modification due to Mesland-Rennie.
For that we have to use unique algebraic data from Section \ref{16sep25sm4}, with the idempotent $\Psi$ from the Mesland-Rennie formalism playing the crucial role in this task. The rationale behind the idempotent $\Psi$ is to identify the quotient space $\Lambda^2_D$ as the image $\left(1 - \Psi\right)T^2_D$, thus avoiding to work with classes of operators. Another main difference of the approach from Mesland--Renie \cite{mesland2025existence} compared to Connes \cite{connes} is that higher order differential calculi are defined using balanced tensor products instead of considering polynomials in represented algebra elements and differentials of represented algebra elements. Finally, we stress that our accomplishment of equating spectral and algebraic torsion functionals would be actually possible also in the approach by Connes if one worked with $\Omega^2_D$ not as orthogonal complement of $J^2_D$ but a complement generated by the special idempotent $(1-\mu\circ\Phi\circ\mu^{-1})$ playing the role of $(1-\Phi)$ in Mesland--Rennie formalism. The tensor constructions would not be necessary for this equality, but achieving it in Mesland--Rennie formalism opens a way to study other geometric tensors besides the torsion.\\

In this respect we compute the various spectral functionals introduced in \cite{Dabrowski:2022ufo,Bochniak:2024duf,Bochniak:2025rud}. Some of them are sensitive to the presence of torsion in the spectral triple. We found that although the Ricci scalar spectral functional is sensitive to torsion, its nonzero contribution involving the Majorana mass matrix $\Upsilon_R$ arises purely from the torsion independent part of curvature, rather
than from the torsion dependent part. As we have shown in the paper, the Dirac operator terms containing the $\Upsilon_R$ term commute with the entire represented algebra $A_F$ and as such, any Mesland--Rennie type of differential calculus, and all of its constructions on it, are blind to this term. Therefore we predict difficulties equating the Ricci scalar spectral functional and the still to be defined algebraic Ricci scalar functional.\\

Finally, we would like to stress that the idempotent $\Psi$ defined in Section \ref{Sec V.B.} can be trivially used to construct a torsionless connection $\nabla$, e.g., by setting $\nabla = 0$ on exact 1--forms (or by adding arbitrary elements of $JT^2_D$ to $\nabla$ acting on exact 1--forms). This $\Psi$ is compatible with the involution and all the assumptions from \cite{mesland2025existence} are completely satisfied. We are not claiming that the results of our Section \ref{16sep25sm4} are in conflict with the unicity results from \cite{mesland2025existence}, since their results are associated with torsion zero connection, and the one crucial for us is definitely not such.
\acknowledgments
\noindent The research of F.P. was supported by Croatian Science Foundation HRZZ's MOBDOK-2023-8742 grant which funded the research visit which resulted in this paper. Additionally, F.P.'s research in his home institution IRB is supported by the Croatian Science Foundation Project No. IP-2025-02-8625, "Quantum aspects of gravity".\\
\noindent S.M. was partially supported with funding from the EU project Caligola (HORIZON-MSCA-2021-SE-01), Project ID: 101086123, and CA21109 - COST Action CaLISTA.\\
\noindent Moreover, this research is part of the EU Staff Exchange project 101086394 "Operator Algebras That One Can See". It was partially supported by the University of Warsaw Thematic Research Programme "Quantum Symmetries".\\

\noindent The authors thank dr.sc. Tajron Juri\'{c} for his comments on the first version of the preprint, and dr.sc. Arkadiusz Bochniak for his comments on the second version of the preprint.

\newpage
\setcounter{section}{0}
\renewcommand{\thesection}{Appendix~\Alph{section}}
\section{\protect\NoCaseChange{The bimodule of junk two forms}} \label{7oct25sm1}

In the definition of the Connes space of two forms \eqref{11sep25sm2}, we had invoked the fact that $J^2_D$ is a bimodule. Here we provide the proof (which seems independent from the proof from Connes's book \cite{ConnesMarcolli2007}, but also from the specifics of the spectral triple that we used in the paper).\\
Let us point out the condition $\sum a_j[D,b_j] = 0$ in the definition of $J^2_D$.
So we have,
\begin{equation}
    \begin{split}
        &\sum [D,c a_j][D,b_j]\\
        =&{[D,c]\sum a_j[D,b_j]} + c \sum [D,a_j][D,b_j]\\
        =&c \sum [D,a_j][D,b_j].
    \end{split}
\end{equation}
The term ${[D,c]\sum a_j[D,b_j]}$ is cancelled due to the prescribed condition. Thus, $J^2_D$ is a left module. The proof of the property of right modularity is slightly more involved.\\
First, we derive the following equation using the Leibniz property to arrive at a form suitable to our definition of $J^2_D$.
\begin{equation}
    0 = \sum a_j [D,b_j] c = \sum a_j [D,b_j c] - \sum a_j b_j [D,c].
\end{equation}
Then, the space $J^2_D$ is a right module of $\pi_D(\Omega^2_u)$ as follows.
\begin{equation}
    \begin{split}
        &\sum [D,a_j][D,b_jc] - \sum [D,a_j b_j][D,c]\\
        =&\sum [D,a_j][D,b_j]c +\sum [D,a_j]b_j[D,c] - \sum [D,a_j b_j][D,c]\\
        =&\sum [D,a_j][D,b_j]c + {\sum [D,a_j b_j][D,c]} - {(\sum a_j[D,b_j])[D,c]} - {\sum [D,a_j b_j][D,c]}\\
        =&\sum [D,a_j][D,b_j]c.
    \end{split}
\end{equation}
Here, we have again invoked the prescribed condition to cancel a term.
\bibliography{BibTex}
\end{document}